\documentclass[usenatbib]{mn2e}
\usepackage{graphicx}
\usepackage{epsfig}
\usepackage{lscape}
\usepackage{longtable}
\usepackage{color}
\usepackage{pdfpages}
\usepackage{epstopdf}
\usepackage{amsmath,amssymb}
\usepackage{multicol}
\usepackage{bm}

\title[LSPs in luminous red giant variables]{Long Secondary Periods in luminous red giant variables}
\author[Masaki Takayama and Yoshifusa Ita]{Masaki Takayama$^{1, 2}$\thanks{E-mail:
takayama@nhao.jp} and Yoshifusa Ita$^{3}$\\
$^{1}$Nishi-Harima Astronomical Observatory, Center for
Astronomy, University of Hyogo, 407-2, Nishigaichi, Sayo-cho, Hyogo
679-5313, Japan\\
$^{2}$Department of Astronomy, School of Science, The University of Tokyo, Bunkyo-ku, Tokyo 113-0033, Japan\\
$^{3}$Astronomical Institute, Graduate School of Science, Tohoku University, Sendai, Miyagi 980-8578, Japan}
\begin{document}

\date{19 Dec 2019}

\pagerange{\pageref{firstpage}--\pageref{lastpage}} \pubyear{}

\maketitle

\label{firstpage}

\begin{abstract}
The origin of  long secondary periods (LSPs) in  red giant variables is unknown.
We investigate whether stellar pulsations in red giants can explain the properties of the LSP variability. 
VIJHKs light curves obtained by OGLE and the IRSF/SIRIUS survey in the Small Magellanic Cloud are examined.
The sample of oxygen-rich LSP stars show evidence of a phase lag between the light curves of optical and near- IR band.
The change in radius contributes the bolometric change roughly half as much as the change in temperature, implying that the change in effective temperature plays an important role in the luminosity change associated with the LSPs.
We have created numerical models based on the spherical harmonics to calculate the light amplitudes of dipole mode variability and have found that the models can roughly reproduce the amplitude - amplitude relations (e.g. ($\Delta I$, $\Delta H$)). 
The LSP variability can be reproduced by the dipole mode oscillations with temperature amplitude of $\lesssim$ 100 K and $\lesssim$ 150 K for oxygen-rich stars and most carbon stars, respectively. 
Radial pulsation models are also examined and can reproduce the observed colour change of the LSPs.
However, there is still an inconsistency in length between the LSP and periods of radial fundamental mode.
On the other hand, theoretical PL relations of the dipole mode corresponding to so-called oscillatory convective mode were roughly consistent with observation. 
Hence our result suggests that the observations can be consistent with stellar pulsations corresponding to oscillatory convective modes.

\end{abstract}


\section{Introduction}
\label{sec:intro}
Recent long-term ground-based observations (e.g. OGLE, MACHO) have developed our knowledge of luminous red giant variables. 
Those stars show light curves with periods longer than 1 day and are, therefore, called long period variables (LPVs).
Recently, at least eight period-luminosity (PL) relations, which are labeled A$^{\prime}$, A, B, C$^{\prime}$, C, D, E and F, have been found among the LPVs in both the Large and Small Magellanic Cloud (LMC, SMC) (\citealt{woo99}; \citealt{sos04}; \citealt{ita04}; \citealt{tab10}; \citealt{sos13b}). 
On the origin of the six sequences, the sequences A$^{\prime}$ - C and F are known as stellar pulsations corresponding to lower order radial and non-radial $p$-modes (\citealt{woo99}; \citealt{ita04}; \citealt{takayama13}; \citealt{sos13b}; \citealt{ste14}; \citealt{woo15}; \citealt{tar17}).
The sequence C and C$^{\prime}$ consist of Mira/Semi-Regular (SR) variables and are interpreted as the radial pulsations corresponding to the fundamental mode (i.e. the radial-pulsation mode corresponding to the longest period) and the first overtone mode of the red giants, respectively (\citealt{ita04}).
On the other hand, LPVs falling in the sequence A$^{\prime}$ - B are classified as the OGLE Small Amplitude Red Giants (OSARGs) (\citealt{sos04}; \citealt{sos07}). 
The sequence E consists of eclipsing binaries and rotating stars with ellipsoidal shape (\citealt{woo99}; \citealt{der06}; \citealt{sos07b}). 

The sequence D stars have periods lying between approximately 400 and 1500 days, which are longer than the period falling on the sequence C in the same luminosity. 
These stars are also pulsating with shorter periods (primary periods) typically falling on the sequence B (\citealt{woo99}).
The periods corresponding to the sequence D are called long secondary periods (LSPs). 
Note that the primary periods do not necessarily correspond to the largest amplitude of the light curves. 
Approximately  25 - 50$\%$ of the luminous red giant variables in MACHO and OGLE database showed the light variations associated with the LSPs (e.g. \citealt{sos07}; \citealt{fra08}; \citealt{nic09}). 
Many explanations for the LSP phenomenon have been proposed, but the origin of the LSPs is still unknown. 
Here is the summary of the points.

Binarity is one of the straightforward explanations to naturally interpret a longer period than those for the radial pulsations.
Hence a close binary system like sequence E stars has been well argued. 
In previous works, many researchers suggested that the sequence E smoothly joined with the bottom of the sequence D on the period-luminosity diagrams if the sequence E stars were plotted using the orbital period (i.e. double the photometric period) (\citealt{sos04_seqE}; \citealt{der06}; \citealt{sos07}). 
\citet{sos07} also found that about 5$\%$ of the LSP stars in the LMC showed eclipsing-like or ellipsoidal-like modulations in their light curves. 
On the other hand, many of the LSP stars found in MACHO database showed light curves with faster decline and slower rise. 
\citet{woo99} suggested that eclipse by a comet-like companion with a gas and dust tail could explain such asymmetric shape of the light curves. 
Although the binary scenario seemed a reasonable explanation for the LSP phenomenon, various observations of radial velocities have conflicted with the binary hypothesis. 
One of the crucial evidence was that the radial-velocity amplitudes related in the LSPs were too small.
The typical value for full velocity amplitudes for the LSP stars in the Galaxy and LMC was about 3.5km s$^{-1}$ (\citealt{woo04}; \citealt{nic09}). 
Nicholls et al. calculated the mass of the hypothetical companion of the LSP star and they derived that typical value for companion mass was about 0.09$M_{\odot}$.  
They estimated that less than 1 $\%$ of low mass main-sequence stars have a companion star with stellar mass lying between 0.06 - 0.12$M_{\odot}$.
This fraction is too small to explain the population of the LSP stars.

The periodic dust formation hypothesis due to mass ejection from the central star has been proposed to the explanation for the LSPs by \citet{woo99}. 
\citet{woo09} found the evidence for mid-IR excess from the photometric data.
This was interpreted as, the LSP stars having a circumstellar disk or patchy clouds consisting of dust.
\citet{takayama15} explored the colour and magnitude variations of LSP stars in the SMC. 
They found that the observation was inconsistent with the models for dust formation along a spherical mass shell though they examined various chemical compositions for the dust models.

A rotating star with star spots has been a potential scenario for the LSP phenomenon. 
Some researchers have proposed the presence of cool spots due to magnetic activity on asymptotic giant branch (AGB) stars (e.g. \citealt{sok99}).
\citet{sos14} found that typical LSP stars in OGLE database showed the light curves similar to those of the rotating stars with dark-spots. 
\citet{oli03} found that the equatorial rotation velocity of LSP stars is typically less than 3 km s$^{-1}$, which implies a rotation period of $\geq$2868 days when the stellar radius of $\sim$170R$_{\odot}$.
However since the LSP is typically less than 1500 d, the upper limit of the rotation velocities is too small to justify the LSP properties.
\citet{takayama15} explored the colour and magnitude variations using the optical and near-IR light curves. 
They found that theoretical models for a cool spot on a rotating star were inconsistent with the observations.

Stellar pulsation is one of the most straightforward explanation to interpret the periodic variability of stars.  
The fundamental modes whose period is the longest among radial pulsation modes, however, correspond to the sequence C of the PL relations (\citealt{ita04}).
Moreover, the LSPs are about 4 times longer than the sequence C periods in a given luminosity.
Hence radial pulsation is ruled out.
On the other hand, non-radial $g$-mode pulsations were proposed to interpret such a long period (\citealt{woo99}).
However, $g^{+}$-mode pulsations are evanescent in convective layer. 
\citet{woo04} suggested that the amplitude expected from the $g^{+}$ modes would be too small to explain the observed light and velocity amplitudes of the LSP stars, since luminous red giant stars have thick convective envelope and very thin outermost radiative layer.

Recently, $g^{-}$-mode pulsations were proposed as the explanation for the LSPs.
In adiabatic condition, the frequencies of $g^{-}$-modes in a star having fully convective envelope are purely imaginary. 
However, in red giants with $\log L/L_{\odot}>$3, non-adiabaticity cannot be ignored and the $g^{-}$-modes become oscillatory in extremely non-adiabatic condition.
These $g^{-}$-modes are so-called oscillatory convective modes (\citealt{saio15}). 
Saio et al. discussed the properties of the oscillatory convective modes in luminous AGB stars. 
They also found that the theoretical PL relations of the dipole mode ($l$=1) roughly agree with the sequence D in the LMC.
However, it has never been examined, whether the non-radial dipole mode oscillations  can reproduce the properties of the light variations associated with the LSPs.

In this paper, we examine a possibility, whether dipole mode oscillations can explain the LSPs.
In Section \ref{sec:model}, our numerical models which calculate the light amplitudes of pulsating stars are discussed.
In Section \ref{sec:Data}, the observational data for the light curves obtained by OGLE (\citealt{sos11}) and IRSF/SIRIUS (\citealt{ita18}) survey in the SMC and the photometric magnitude data obtained by the Magellanic Clouds Photometric Survey (MCPS) catalog (\citealt{zar02}) and the {\it Spitzer} SAGE SMC IRAC (Surveying the Agents of Galaxy Evolution SMC Infrared Array Camera) source catalog (\citealt{gor11}) are introduced.
In Section \ref{sec:ana}, the observation data are analysed and in Section \ref{sec:result}, the properties of the LSP variability are discussed.
In Section \ref{sec:discs}, our models are compared. 
We discuss a possibility of dipole mode variability as explanation for LSPs.

\section{The models}
\label{sec:model}
We made models to calculate the brightness and the light amplitudes of a non-radially pulsating star.
The model light amplitudes vary with the inclination angle $\alpha$ between the pulsation axis and the line of sight due to the non-spherical distribution of the light intensity on the photosphere.
We modelled a star using the spherical harmonics $Y^{0}_{1}$ (the dipole modes) for the temperature distributions in the photospheres and examined models with various $\alpha$.

The brightness of the star is calculated by integration of the intensity of the brightness along the stellar disc.  
For the radiation flux from the photosphere, we utilised spherically symmetric MARCS code (\citealt{gus08}) with chemical compositions of "{\it heavily CN-cycled}"  and [Fe/H]=-1.0 for oxygen-rich stars and blackbody properties for carbon stars, respectively.
We also assume a stellar mass of 1.0M$_{\odot}$ and surface gravity of $\log g$=0.5 for the models.

In our models, we assume the mean temperature of the photosphere ($T_{0}$) to be $T_{0}=3700$ K and $T_{0}=3200$ K for oxygen-rich star and carbon star, respectively.
As mentioned in Sec \ref{T-vari}, these values are consistent with the typical values for the mean effective temperatures of the oxygen-rich stars and carbon stars, respectively.

Our models also depend on the temperature amplitudes ($\Delta T$) and the inclination angle ($\alpha$).  
We examine models for $\alpha$=0, 30, 60, and 90 degrees with $\Delta T$=100 K, 150 K and 200 K.
Note that $\alpha=0$ corresponds to when the pulsation axis agrees with the line of sight direction.
To consider the dimming of light from the star due to the limb darkening effect, both  models with and without limb darkening have been calculated.
We used the linear limb darkening coefficients of \citet{nei13}.
The full light amplitudes in various wavebands are calculated for models with a given inclination angle and temperature amplitude.
For more detail of the models, see Appendix \ref{apA}.

\section{The observational data}
\label{sec:Data}
\subsection{The near-infrared data}
\label{sec:NIRdata}
Long-term magnitude variations in three near-infrared bands ($J(1.25\mu m)$, $H(1.63\mu m)$, and $K_{\rm s}(2.14\mu m)$) are available for a number of variable stars in the LMC/SMC by the observations at the South African Astronomical Observatory at Sutherland with IRSF/SIRIUS camera (\citealt{ita18}). 
The observations with SIRIUS camera have operated within 1$^\circ$ $\times$ 1$^\circ$ area around the centre in the SMC more than one hundred times  from July 2001 to Dec 2017, and a total of $\sim$61,000 variable source candidates were detected (for more detail, see \citealt{ita18}).
Among those candidates, 25,119 stars have time series data of all the three near-infrared bands.
In this work, we picked up the LSP candidates proposed by \citet{sos11} among the 25,119 stars, and used the time series data.

\subsection{The optical data}
We obtained the $V$ and $I$ band light curves of red giant variables in the SMC from the OGLE database (\citealt{sos11}).
OGLE -II and -III observed huge number of stars in the SMC about 50-100 times with the $V$ band filter and about 700-1,000 times with the $I$ band filter, respectively, from 1997 to 2009.
\citet{sos11} obtained LSP candidates from the OGLE database according to the position on the PL diagram. 
They also divided the red giants into oxygen-rich stars and carbon stars. 
We found 388 LSP candidates consisting of 300 oxygen rich stars and 88 carbon stars in the area of the IRSF/SIRIUS survey.

\subsection{The $U$, $B$ and mid-IR band data}
In addition to the $VIJHK_{\rm s}$ data of OGLE and the IRSF/SIRIUS projects, we use the Magellanic Clouds Photometric Survey (MCPS) catalog (\citealt{zar02}) for the $U$ and $B$ band magnitudes and the {\it Spitzer} SAGE SMC IRAC (Surveying the Agents of Galaxy Evolution SMC Infrared Array Camera) source catalog (\citealt{gor11}) for the mid-IR ([3.6][4.5][5.8][8.0]) band magnitudes.
We consider the individual magnitude values reported in these catalogs as mean magnitudes of each star.

\section{data analysis}
\label{sec:ana}
In this section, we obtain the light curve parameters for the optical and near-IR bands by using the least square periodgram and select a sample of LSP stars.
Then, by using the photometric magnitudes of various wavebands, we obtain the time series of the effective temperature, bolometric luminosity, and stellar radius of the stars. 

\subsection{sample selection}
\label{select}

To investigate the properties of the light variations associated with the LSP, we selected a sample of LSP stars by combining the OGLE and IRSF/SIRIUS database. 
From the 388 LSP candidates found in the area of the SIRIUS survey, we chose samples that show periodicities corresponding to the LSP in both optical and NIR light curves.

The periods corresponding to the light curves of the star are determined by the least-square periodogram based on the Lomb-Scargle periodogram (\citealt{lom76}; \citealt{sca82}).  
The brief explanation of our method is as follows;
We consider a first-order Fourier series
\begin{eqnarray}
m_{P}(t_{i})&=&m_{0}+A_{P} \sin \left( 2 \pi \frac{t_{i}-t_{0}}{P} +2\pi \phi_{P} \right) \nonumber \\
& & \:\:\:\:\:\:\:\:\:\:\:\: \:\:\:\:\:\:\:\:\:\:\:\: \:\:\:\:\:\:\:\:\:\:\:\:(-0.5\leq \phi < 0.5),
\label{sin}
\end{eqnarray}
where $t_{i}$ is the observation time ($i$=1,2,3,,,) while $m_{0}$ and $A_{P}$ are the offset term and the amplitude for a given $P$ value, respectively. 
The initial phase $\phi_{P}$ is determined with a reference time $t_{0}$ corresponding to JD 2,450,000. 
A chi-squared distribution $\chi^{2}$ against $P$ value is given by
\begin{eqnarray}
\chi^{2}(P) \equiv \sum_{i} \left(m(t_{i})-m_{P}(t_{i}) \right)^{2},
\label{eq_chi}
\end{eqnarray}
where $m_{i}$ is the observed magnitude in $t_{i}$.
We searched a $P$ value corresponding to minimal value for $\chi^{2}$ in the range of 1.01 - $t_{\rm max}$ d in a step of 0.01 d, where $t_{\rm max}$ is a half the length of the observation period for the star. 

\begin{figure}
\includegraphics[width=0.5\textwidth]{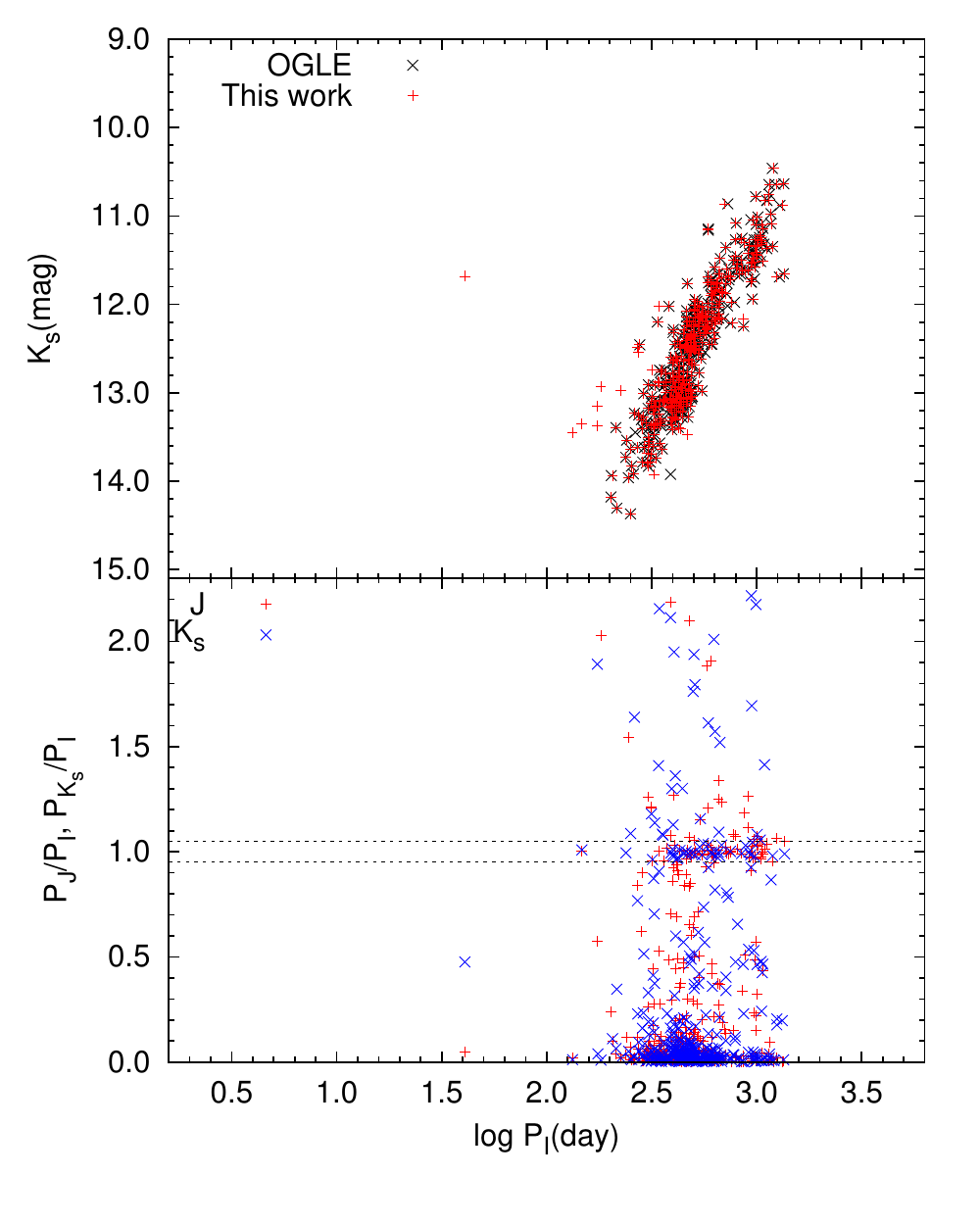}
\caption{The top panel shows the period-$K_{\rm s}$ magnitude relation consisting of a sample of 388 LSP candidates. 
The red pluses correspond to the period $P_{I}$ derived by our method, while the black crosses are the period published by OGLE database (\citealt{sos11}). 
The root mean square (RMS) of the difference between $P_{I}$ and $P_{\rm OGLE}$ (i.e. $(P_{I}-P_{\rm OGLE})/P_{\rm OGLE}$) for 388 LSP candidates was 2.6$\%$.
$K_{\rm s}$ magnitudes correspond to the mean magnitude of the stars obtained from SIRIUS data.
The bottom panel shows a plot of the period ratios in those stars.
The red pulses and the blue crosses are the $P_{J}/P_{I}$ and $P_{K_{\rm s}}/P_{I}$ values, respectively. 
The two black dashed horizontal lines correspond to period ratios of 0.95 and 1.05.
}
\label{PL_Pratio}
\end{figure}

Our method for period analysis obtain a period corresponding to the mode of the largest amplitude in the light curves.
We obtained period $P_{I}$ from $I$ light curves of 388 LSP candidates by using our method.
The top panel of Figure\,\ref{PL_Pratio} shows the plots on $P_{I}$-$K_{s}$ diagram.
We confirmed that the obtained PL relation is consistent with the sequence D found by previous works (\citealt{sos11}).
Hence we consider that the $P_{I}$ is consistent with the LSP of the star.  

We also obtained periods $P_{J}$ and $P_{K_{\rm s}}$ from $J$ and $K_{\rm s}$ light curves.
The bottom panel of Figure\,\ref{PL_Pratio} shows the period ratios $P_{J}/P_{\rm I}$ and $P_{K_{s}}/P_{\rm I}$, respectively. 
If the mode corresponding to the LSP is prominent in not only $I$ light curves but also the NIR light curves, both $P_{J}$ and $P_{K_{\rm s}}$ have similar values to the LSP in the $I$ light curves (i.e. $P_{I}$).
Hence we selected a sample of the stars whose $P_{J}$ and $P_{K_{\rm s}}$ values both fall within 5$\%$ of the $P_{I}$ value. 
As a result of this selection, we ended up with a sample of 18 LSPs consisting of 9 oxygen-rich stars and 9 carbon stars. 
We also calculated the best fit values for light amplitude, initial phase, and offset term for $V$, $J$, $H$, and $K_{\rm s}$ light curves by adopting $P_{I}$.
Table~\ref{tab01} shows the best fit values for the light curve parameters.

\begin{table*}
\caption{Light curve parameters.}
\scalebox{0.8}[0.8]{
\begin{tabular}[l]{cccccccccccccccccc}
\hline

ID$^{a}$ & Sp & $P_{I}$ & $V$ & $\Delta V^{b}$ & $\phi_{V}$ & $I$ & $\Delta I^{b}$ & $\phi_{I}$ & $J$ & $\Delta J^{b}$ & $\phi_{J}$ & $H$ & $\Delta H^{b}$ & $\phi_{H}$ & $K_{\rm s}$ & $\Delta K_{\rm s}^{b}$ & $\phi_{K_{\rm s}}$\\
 & type & {\rm (day)} & {\rm (mag)} & {\rm (mag)} &  & {\rm (mag)} & {\rm (mag)} & & {\rm (mag)} & {\rm (mag)} & & {\rm (mag)} & {\rm (mag)} & & {\rm (mag)} & {\rm (mag)} &  \\
\hline
\hline
07522 & O-rich & 319.06 & 17.24 & 0.43 & 0.049 & 15.15 & 0.16 & 0.004 & 13.92 & 0.14 & -0.055 & 13.07 & 0.13 & -0.024 & 12.73 & 0.09 & -0.021 \\ 
07563 & O-rich & 548.97 & 16.61 & 0.13 & -0.235 & 14.60 & 0.08 & -0.240 & 13.31 & 0.05 & -0.284 & 12.43 & 0.04 & -0.180 & 12.16 & 0.05 & -0.264 \\ 
07849 & O-rich & 431.55 & 16.76 & 0.41 & 0.269 & 14.78 & 0.18 & 0.190 & 13.47 & 0.05 & 0.168 & 12.63 & 0.04 & 0.145 & 12.45 & 0.06 & 0.198 \\ 
07852 & O-rich & 641.00 & 16.67 & 0.46 & -0.296 & 14.37 & 0.22 & -0.266 & 12.93 & 0.05 & -0.274 & 12.09 & 0.04 & -0.289 & 11.81 & 0.04 & -0.215 \\ 
07858 & O-rich & 509.19 & 16.84 & 0.19 & -0.340 & 14.99 & 0.10 & -0.309 & 13.63 & 0.08 & -0.374 & 12.84 & 0.06 & -0.312 & 12.51 & 0.05 & -0.247 \\ 
08978 & O-rich & 418.78 & 16.67 & 0.37 & 0.012 & 14.75 & 0.16 & 0.043 & 13.46 & 0.02 & -0.085 & 12.63 & 0.03 & 0.030 & 12.42 & 0.04 & 0.098 \\ 
09444 & O-rich & 146.88 & 14.03 & 0.10 & 0.322 & 13.71 & 0.08 & 0.353 & 13.55 & 0.04 & 0.281 & 13.41 & 0.06 & 0.318 & 13.35 & 0.03 & 0.385 \\ 
09856 & O-rich & 613.89 & 16.58 & 0.27 & 0.041 & 14.40 & 0.22 & -0.037 & 13.01 & 0.05 & -0.065 & 12.18 & 0.05 & 0.061 & 11.88 & 0.06 & 0.092 \\ 
13234 & O-rich & 489.95 & 16.66 & 0.26 & -0.206 & 14.69 & 0.16 & -0.232 & 13.34 & 0.06 & -0.189 & 12.51 & 0.05 & -0.220 & 12.25 & 0.06 & -0.184 \\ 
07622 & C-rich & 1190.29 & 17.00 & 0.40 & -0.121 & 14.44 & 0.18 & -0.033 & 12.80 & 0.10 & -0.038 & 11.86 & 0.12 & -0.009 & 11.35 & 0.08 & -0.010 \\ 
08199 & C-rich & 749.95 & 18.27 & 1.04 & 0.296 & 15.26 & 0.59 & 0.290 & 13.46 & 0.41 & 0.217 & 12.43 & 0.33 & 0.240 & 11.73 & 0.21 & 0.243 \\ 
09688 & C-rich & 588.32 & 16.69 & 0.12 & -0.187 & 14.54 & 0.14 & -0.183 & 13.04 & 0.07 & -0.292 & 12.22 & 0.09 & -0.247 & 11.68 & 0.06 & -0.296 \\ 
10109 & C-rich & 977.93 & 16.20 & 0.50 & 0.421 & 13.98 & 0.24 & 0.455 & 12.58 & 0.16 & 0.442 & 11.70 & 0.14 & 0.495 & 11.27 & 0.07 & -0.484 \\ 
10309 & C-rich & 952.61 & 16.49 & 0.18 & 0.085 & 14.29 & 0.17 & 0.109 & 12.78 & 0.15 & 0.090 & 11.92 & 0.11 & 0.103 & 11.53 & 0.06 & 0.049 \\ 
11688 & C-rich & 846.77 & 16.20 & 0.21 & 0.296 & 14.20 & 0.13 & 0.344 & 12.66 & 0.07 & 0.378 & 11.86 & 0.04 & 0.433 & 11.50 & 0.03 & 0.384 \\ 
12653 & C-rich & 1356.29 & 17.70 & 0.40 & 0.430 & 14.90 & 0.30 & 0.407 & 13.21 & 0.08 & 0.423 & 12.19 & 0.14 & 0.375 & 11.64 & 0.08 & 0.386 \\ 
13748 & C-rich & 936.41 & 16.99 & 0.44 & 0.276 & 14.87 & 0.36 & 0.336 & 13.18 & 0.10 & 0.361 & 12.22 & 0.13 & 0.370 & 11.74 & 0.07 & 0.357 \\ 
13945 & C-rich & 918.45 & 16.49 & 0.30 & 0.098 & 14.18 & 0.21 & 0.092 & 12.72 & 0.10 & 0.082 & 11.82 & 0.11 & 0.109 & 11.34 & 0.08 & 0.106 \\ 
\hline
\multicolumn{3}{l}{$^{a}$OGLE-SMC-LPV-xxxxx} \\
\multicolumn{3}{l}{$^{b}$full amplitude} \\
\end{tabular}
}
\label{tab01}
\end{table*}

\subsection{The combined optical, and near- and mid-IR data}
We wish to combine the optical and near-IR data and obtain the $VIJHK_{\rm s}$ magnitudes at the same observation time.
However, there was no SIRIUS data at the same observation time as OGLE data.
In addition, the observations with $I$ filter were more frequent than the observations with the $JHK_{\rm s}$ filters.
Hence we calculated the $I$ band magnitudes at the observation time of SIRIUS ($t_{\rm SIRIUS}$) by using linear interpolation between $I(t_{i})$ and $I(t_{i+1})$, where  $t_{i} \le t_{\rm SIRIUS} \le t_{i+1}$.
Note the interpolations were only used for when $t_{i+1}-t_{i} \le$ 5 d.

\begin{figure}
\includegraphics[width=0.5\textwidth]{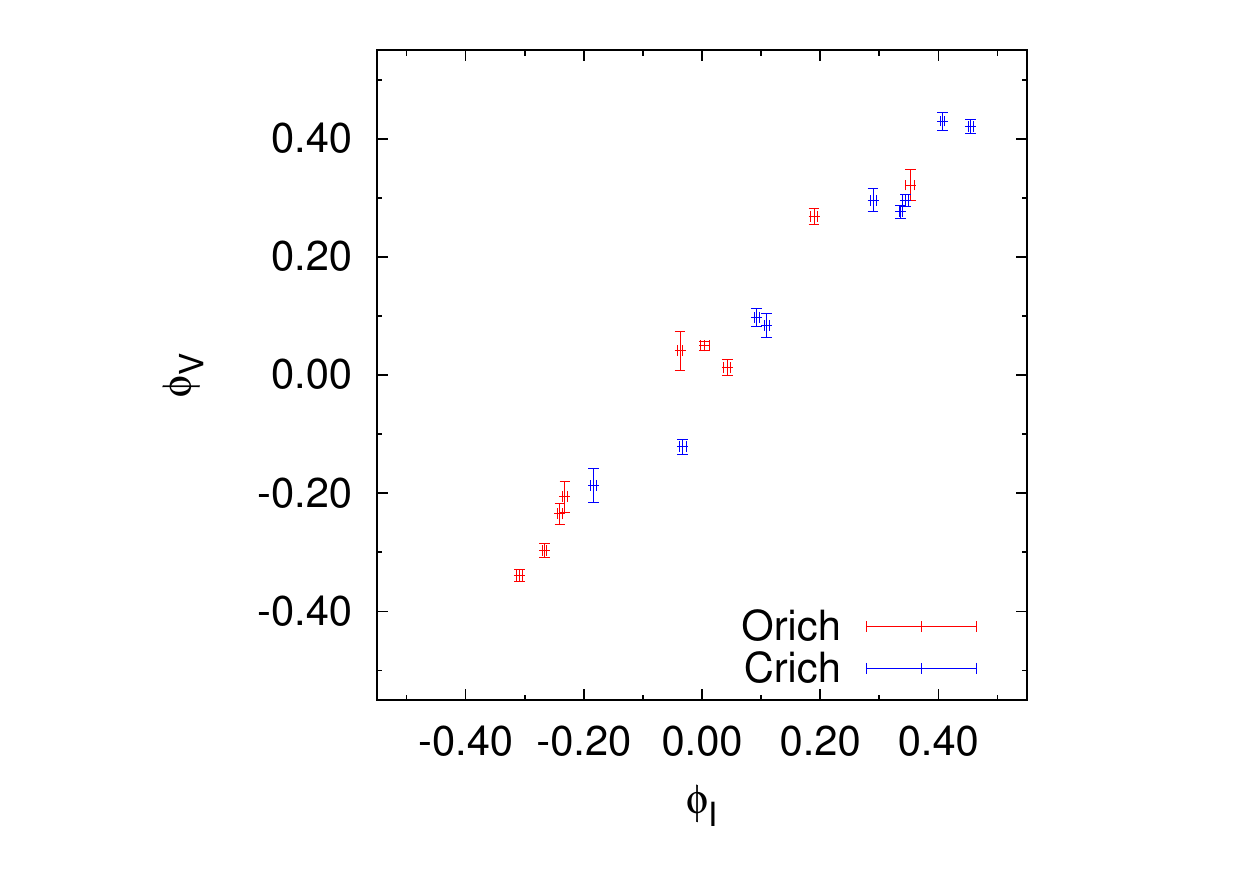}
\caption{Comparison of the initial phase for $I$ band against these for $V$ band. 
The red and blue colours correspond to the oxygen-rich and carbon stars, respectively. 
The error bars correspond to 1$\sigma$ uncertainty, where $\sigma$ is the error of a fit. 
The amount of the phase shift obtained by a fit with a linear function $\phi_{V}=\phi_{I} + a$ where the offset was calculated to be $a$=0.00. 
Therefore there is little to no evidence for phase lag between the $I$ light curves and $V$ light curves.}
\label{phase_IV}
\end{figure}

The time series of the $V$ band were too sparse to use our method for the $I$ band data.
However, the light curves associated with the LSPs were almost coherent between the $I$ band and $V$ band (Figure\,\ref{phase_IV}).
Hence we obtained the $V$ band light curves of the star by using the following equation,
\begin{eqnarray}
V(t_{\rm SIRIUS})=\frac{\Delta V}{\Delta I}(I(t_{\rm SIRIUS})-\left < I \right > )+ \left < V \right >,
\label{Vt}
\end{eqnarray}
where $\Delta V$ and $\Delta I$ correspond to the amplitudes of $V$ and $I$ bands, respectively. 
$\left< V \right>$ and $\left< I \right>$ are the mean magnitudes of $V$ and $I$ bands, respectively. 

\begin{figure}
\includegraphics[width=0.5\textwidth]{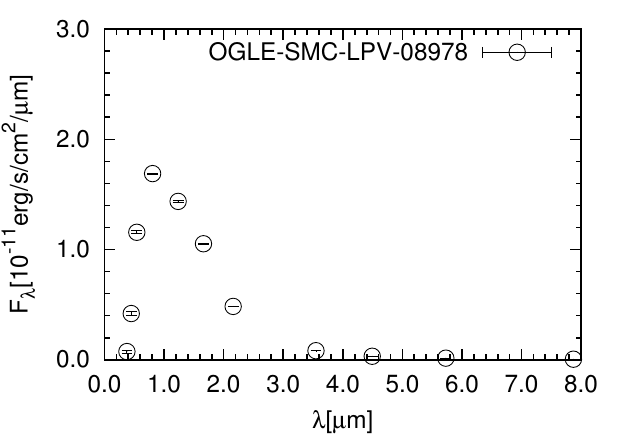}
\includegraphics[width=0.5\textwidth]{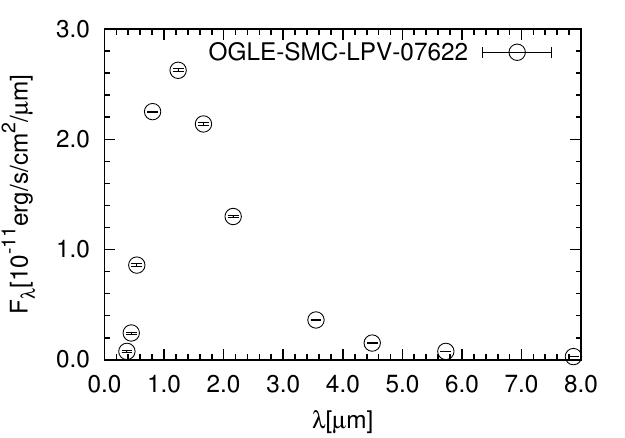}
\caption{Examples of the spectral energy distributions of our program stars. The upper panel corresponds to OGLE-SMC-LPV-08978 (oxygen-rich star) while the bottom panel is of OGLE-SMC-LPV-07622 (carbon star). The zero magnitude flux and its reference wavelength are referred from \citet{coh03a} for $UBVI$ bands, \citet{coh03b} for $JHK_{\rm s}$ bands, \citet{irac06} for [3.6]-[8.0] bands. The error bars correspond to 1$\sigma$ uncertainty of the observation error}
\label{SED}
\end{figure}

In order to compute apparent bolometric luminosities, we require
a spectral energy distribution which we obtain from $UBVIJHK_{\rm s}$, [3.6], [4.5], [5.8], and [8.0] band magnitudes. 
Due to lack of observed amplitude data, we assume the $U, B$ and the mid-IR magnitudes to be constant during the LSP cycles.
Figure\, \ref{SED} shows examples of the spectral energy distributions (SEDs) of the stars.
The energy flux emitted by red giants such as LSPs is confined mainly between the $V$ and $K_{\rm s}$ bands.
It implies that most of the amount of fluctuation of energy flux would result from variations in this wavelength range.
Hence it is adequate to assume a constant magnitude for the $UB$ and mid-IR bands as we compute the perturbations of the bolometric luminosity.

For the correction of the reddening due to the interstellar extinction in our Galaxy and the SMC, the colour excess $E(B-V)$ of 0.12 mag (\citealt{kel06}) and the mean $R_{\rm V}$-dependent extinction law of \citet{car89} with $R_{\rm V}$=3.2 are used. 
By using the photometric magnitudes of various waveband, we obtain the time series of the effective temperature, bolometric luminosity, and stellar radius of the stars and explore how those parameters change with the LSP cycle, as explained below.

\subsection{The effective temperatures}
We estimated the effective temperatures of the star by using colour-temperature relations for AGB stars proposed by previous works. 
There is room for debate whether a change in the colour of the central star are caused by a change in the effective temperature since the origin of the light variations of the LSPs is unknown.
Nevertheless, we assumed that the colours of the star always reflect the effective temperature.

The time series for the light variations are used to obtain the time series for the colours. 
However, the $V$ light curves are not used for this purpose because the estimation using Equation \ref{Vt} would involve a significant error for the $V$ magnitude.
\citet{hou00} obtained the $(V-I)-T_{\rm eff}$ and $(V-K)-T_{\rm eff}$ relations for oxygen-rich AGB stars and we derived the $(I-K)-T_{\rm eff}$ relations from those relations.
For the oxygen-rich stars, we obtained the effective temperatures by applying the $(I-K)-T_{\rm eff}$ relations corresponding to [Fe/H] of -1.0 (as appropriate for the SMC) and $\log g$ of 0.5 to the $I-K_{\rm s}$ magnitudes. 
While for the carbon stars, the $(J-K)-T_{\rm eff}$ relation (\citealt{bes83}) was used.
By using the least square fit with a single sine curve (given by Equation \ref{sin}) with $P_{I}$ to the time series of $T_{\rm eff}$, the amplitude $\Delta T$ and mean values $\left< T \right>$ were derived (Table~\ref{tab02}).
Note that the errors for $T_{\rm eff}$ have been derived from the photometric magnitudes and its errors.  We did not consider the uncertainty of the colour-$T_{\rm eff}$ relations.

\subsection{The bolometric luminosities}
The apparent bolometric luminosities were obtained by an integration of the energy flux corresponding to the $UBVIJHK_{\rm s}$, [3.6], [4.5], [5.8], and [8.0] band magnitudes.  
The distance modulus of 18.96 mag for the SMC (\citealt{sco16}) was used for computing the absolute bolometric luminosities. 
Then, by using Equation \ref{sin} for a fit to the time series of the bolometric luminosity, the amplitude $\Delta L$ and mean luminosity $\left< L \right>$ were derived (Table~\ref{tab02}).

\subsection{The stellar radii}
We obtained the time series of the effective temperatures and the bolometric luminosities of the stars from the data.
Then, we computed the stellar radius and obtained its time series by using the Stefan-Boltzmann law,  
\begin{eqnarray}
L=4\pi \sigma R^{2} T_{\rm eff}^{4}.
\label{SBL}
\end{eqnarray}
As for $T_{\rm eff}$ and $L$, the amplitudes $\Delta R$ and mean radii $\left< R \right>$ were obtained by least square fitting of Equation \ref{sin} (Table~\ref{tab02}).

\begin{table*}
\caption{Stellar parameters and their amplitudes} 
\begin{tabular}{ccccccc}
\hline
ID$^{a}$ & $T$ & $\Delta T^{b}$ & $L$ & $\Delta L^{b}$ & $R$ & $\Delta R^{b}$\\
 & (K) & (K) & $(L_{\odot})$ & $(L_{\odot})$ & $(R_{\odot})$ & $(R_{\odot})$\\
\hline
07522 & 3742.4 $\pm$ 9.0 & 54.3 $\pm$ 12.0 & 2134.4 $\pm$ 19.3 & 156.0 $\pm$ 25.3 & 110.0 $\pm$ 0.5 & 1.1 $\pm$ 0.6 \\ 
07563 & 3731.2 $\pm$ 5.0 & 23.6 $\pm$ 7.4 & 3616.2 $\pm$ 9.4 & 88.0 $\pm$ 12.8 & 144.2 $\pm$ 0.4 & 0.7 $\pm$ 0.7 \\ 
07849 & 3800.3 $\pm$ 4.5 & 54.8 $\pm$ 6.3 & 3010.5 $\pm$ 9.2 & 137.0 $\pm$ 12.7 & 126.8 $\pm$ 0.2 & 1.0 $\pm$ 0.3 \\ 
07852 & 3666.8 $\pm$ 4.5 & 68.4 $\pm$ 6.0 & 4808.2 $\pm$ 16.3 & 236.0 $\pm$ 22.2 & 172.1 $\pm$ 0.4 & 2.2 $\pm$ 0.5 \\ 
07858 & 3700.2 $\pm$ 4.7 & 18.6 $\pm$ 6.5 & 2596.1 $\pm$ 7.9 & 92.1 $\pm$ 10.1 & 124.2 $\pm$ 0.4 & 1.4 $\pm$ 0.6 \\ 
08978 & 3802.1 $\pm$ 9.5 & 41.4 $\pm$ 15.7 & 3072.1 $\pm$ 25.2 & 142.0 $\pm$ 38.0 & 127.9 $\pm$ 0.7 & 1.2 $\pm$ 0.9 \\ 
09444 & 5480.0 $\pm$ 6.4 & 29.0 $\pm$ 9.2 & 8470.2 $\pm$ 30.3 & 235.0 $\pm$ 37.9 & 102.3 $\pm$ 0.2 & 0.6 $\pm$ 0.2 \\ 
09856 & 3663.5 $\pm$ 4.8 & 59.2 $\pm$ 6.5 & 4519.3 $\pm$ 13.4 & 180.0 $\pm$ 18.1 & 167.2 $\pm$ 0.4 & 2.1 $\pm$ 0.6 \\ 
13234 & 3716.7 $\pm$ 3.8 & 50.2 $\pm$ 5.3 & 3351.1 $\pm$ 10.2 & 156.0 $\pm$ 14.0 & 139.8 $\pm$ 0.3 & 0.6 $\pm$ 0.4 \\ 
07622 & 3102.3 $\pm$ 13.4 & 32.5 $\pm$ 19.7 & 6128.8 $\pm$ 35.9 & 322.0 $\pm$ 51.3 & 272.2 $\pm$ 2.4 & 4.3 $\pm$ 3.1 \\ 
08199 & 2787.2 $\pm$ 10.2 & 117.0 $\pm$ 23.2 & 4124.2 $\pm$ 25.1 & 464.0 $\pm$ 51.2 & 276.2 $\pm$ 1.9 & 9.3 $\pm$ 4.5 \\ 
09688 & 3245.8 $\pm$ 6.9 & 8.6 $\pm$ 9.7 & 4859.3 $\pm$ 12.7 & 167.0 $\pm$ 18.0 & 221.0 $\pm$ 1.0 & 2.6 $\pm$ 1.4 \\ 
10109 & 3341.8 $\pm$ 11.7 & 76.6 $\pm$ 16.6 & 7443.0 $\pm$ 37.7 & 453.0 $\pm$ 50.1 & 258.2 $\pm$ 1.8 & 6.1 $\pm$ 2.7 \\ 
10309 & 3387.6 $\pm$ 14.8 & 98.1 $\pm$ 21.5 & 5757.3 $\pm$ 24.2 & 380.0 $\pm$ 35.2 & 221.4 $\pm$ 1.8 & 6.0 $\pm$ 2.7 \\ 
11688 & 3557.6 $\pm$ 6.8 & 27.5 $\pm$ 10.2 & 6184.6 $\pm$ 13.4 & 140.0 $\pm$ 17.4 & 207.5 $\pm$ 0.7 & 2.4 $\pm$ 1.1 \\ 
12653 & 2982.5 $\pm$ 38.4 & 65.2 $\pm$ 55.9 & 4389.8 $\pm$ 46.3 & 182.0 $\pm$ 54.4 & 249.2 $\pm$ 6.4 & 9.9 $\pm$ 11.8 \\ 
13748 & 3130.1 $\pm$ 9.0 & 9.3 $\pm$ 12.1 & 4269.3 $\pm$ 14.5 & 273.0 $\pm$ 19.5 & 222.8 $\pm$ 1.2 & 5.8 $\pm$ 1.7 \\ 
13945 & 3221.9 $\pm$ 14.4 & 19.2 $\pm$ 20.3 & 6431.8 $\pm$ 27.0 & 314.0 $\pm$ 38.6 & 258.7 $\pm$ 2.4 & 5.6 $\pm$ 3.3 \\ 
\hline
\multicolumn{2}{l}{$^{a}$OGLE-SMC-LPV-xxxxx} \\
\multicolumn{2}{l}{$^{b}$half amplitude} \\
\end{tabular}
\label{tab02}
\end{table*}

\section{results}
\label{sec:result}
\subsection{The phase lag between the optical and near-IR light curves}
\label{phaseshift}
\begin{figure}
\includegraphics[width=0.47\textwidth]{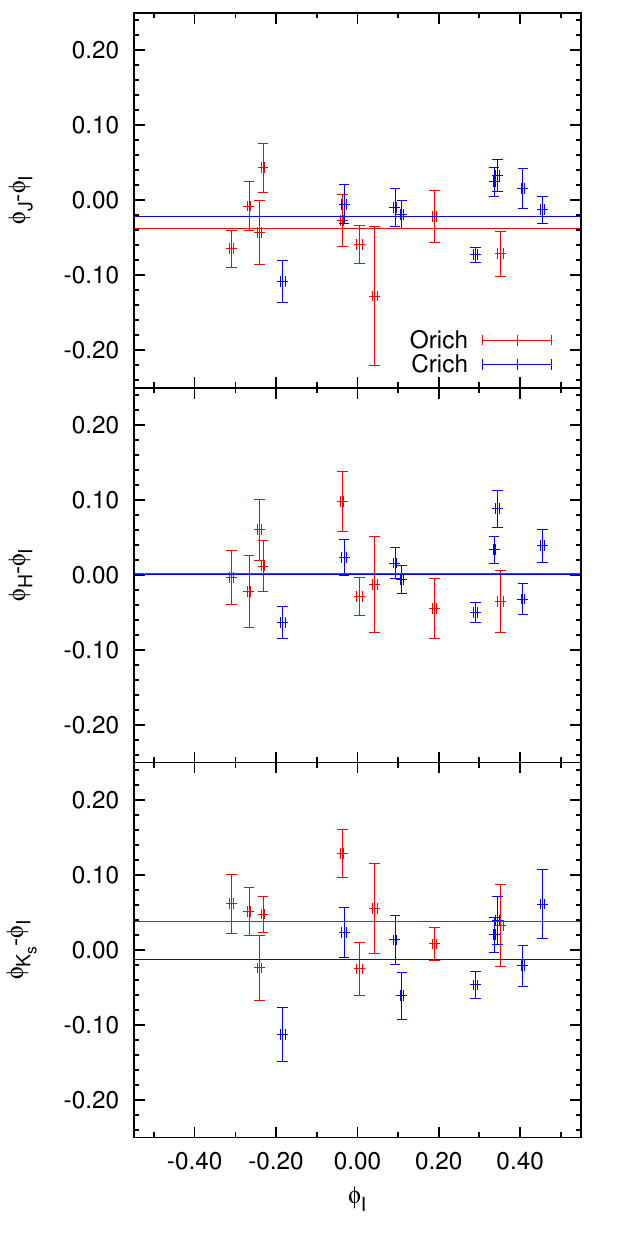}
\caption{Phase lag between the optical and near-IR light curves. The red and blue solid lines  correspond to the mean value for the phase lags for oxygen-rich and the carbon stars, respectively.}
\label{phase}
\end{figure}

The phase lags of the light curves between optical and infrared bands for Mira type variables, a typical long period variable in red giants, have been well investigated (e.g. \citealt{smi02}).
The development of strong molecular absorption (e.g. titanium oxide (TiO) and vanadium oxide (VO) for oxygen-rich stars, and C$_{2}$ and CN for carbon stars) in different layers of the stellar atmospheres at different pulsation phase has been proposed to explain the phase lags.
On the other hand, little is known about properties of phase lags among the light curves associated with the LSPs.

We investigate phase lags given by $\phi_{\rm NIR} - \phi_{I}$. 
Three panels of Figure\,\ref{phase} show plots of the phase lags against $\phi_{I}$. 
The weighted mean values and its standard errors for the phase lags for the oxygen-rich stars are $\phi_{J} - \phi_{I}=-0.038 \pm 0.013$, $\phi_{H} - \phi_{I}=0.002 \pm 0.015$, and $\phi_{K_{\rm s}} - \phi_{I}=0.037 \pm 0.015$, respectively, while for the carbon stars are $\phi_{J} - \phi_{I}=-0.022 \pm 0.015$, $\phi_{H} - \phi_{I}=0.001 \pm 0.016$, and $\phi_{K_{\rm s}} - \phi_{I}=-0.013 \pm 0.016$, respectively.
In the oxygen-rich stars, the mean phase lag between the $I$ and $H$ bands corresponds to zero within 1$\sigma$ uncertainty. 
On the other hand, the mean values for $\phi_{J} - \phi_{I}$ and $\phi_{K_{\rm s}} - \phi_{I}$ significantly differ from zero (more than 3$\sigma$ and 2$\sigma$ for $\phi_{J} - \phi_{I}$ and $\phi_{K_{\rm s}} - \phi_{I}$, respectively).
This suggests that there is a statistically significant amount of the phase lags between the pairs of the $I$ and $J$ bands, and $I$ and $K_{\rm s}$ bands.
From the sign of the mean phase lags, the $J$ band phase follows the $I$ band phase while $K_{\rm s}$ band phase precedes it.
In contrast, the carbon stars have a statistically small amount of phase lag in both the $\phi_{H} - \phi_{I}$ and $\phi_{K_{\rm s}} - \phi_{I}$.
This means that there are not significant phase lags between $I$, $H$, and $K_{\rm}$ bands.
The mean value of the $\phi_{J} - \phi_{I}$ differs from zero by a factor of only $\sim$1.4 of the standard error $\sigma$.
More observational data would be necessary to argue whether the phase lag between the $I$ and $J$ bands is zero or non-zero. 

Although the origin of the LSPs is still unknown, our results suggests a possibility of the strong molecular absorption in the stellar atmosphere.
The amount of the phase lags in the oxygen-rich stars differs from these in the carbon stars, implying difference in molecules associated with the band absorption.

\subsection{The effective temperature variations}
\label{T-vari}
\begin{figure}
\includegraphics[width=0.5\textwidth]{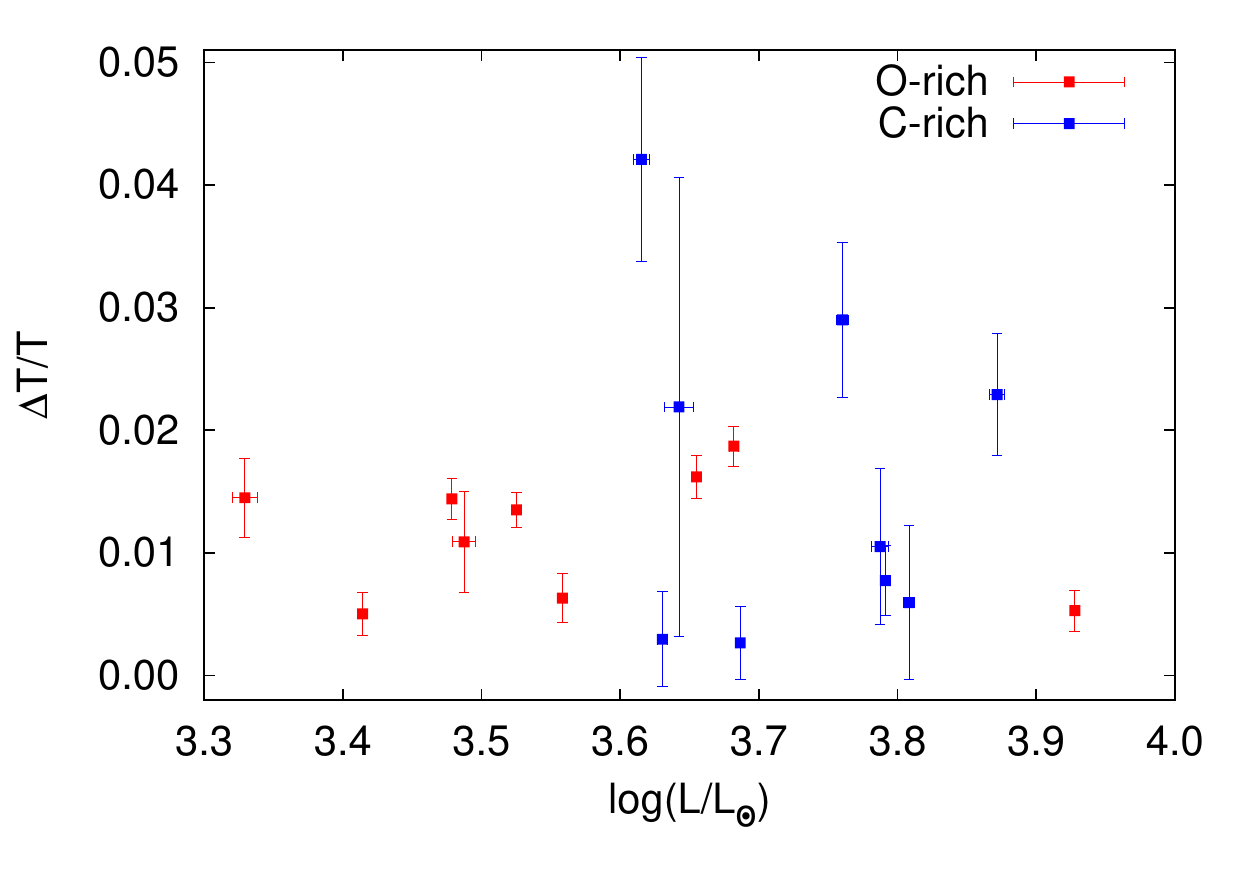}
\caption{Effective-temperature amplitude plotted against mean luminosity in solar unit.}
\label{L_dT}
\end{figure}

The median value for the mean effective temperature $\left< T \right>$ for the oxygen-rich stars was found to be about 3730 K.
A similar value was found in the typical value for the oxygen-rich LSP stars in the LMC (cf. \citealt{nic09})
On the other hand, the value for the carbon stars was $T_{\rm eff} \sim$ 3220 K.
Figure\, \ref{L_dT} shows plots with our program stars for the relation between the mean luminosity ($\log (L /L_{\odot})$) and the effective-temperature amplitude ($\Delta T /T$).
Most stars have temperature amplitudes lying between 10 K and 100 K corresponding to $\Delta T/T$ of 0.003 - 0.04 and the median value for entire samples is about 50 K.
This result is consistent with those for LSP stars in the LMC (cf. \citealt{nic09}).

According to the $I-K_{\rm s}$, $T_{\rm eff}$ and $J-K_{\rm s}$, $T_{\rm eff}$ relations for oxygen-rich and carbon stars, respectively, most stars have mean effective temperatures lying between 2800 and 3800 K.
However, OGLE-SMC-LPV-09444 (classified into the oxygen-rich star) has $\left< T \right>$=5480 K, which is much higher than the typical value for red giants.
The $V-I$ and $J-K_{\rm s}$ are about 0.31 and 0.20, respectively, being consistent with the literatures (cf. \citealt{men02}; \citealt{cut03}). 
Moreover, this star had a similar light curve when compared to the Galactic Be stars (\citealt{cut03}).
Hence OGLE-SMC-LPV-09444 might be different from typical LSP stars in the SMC. Therefore further observation might be necessary.

\subsection{The bolometric variations}
\label{bolvari}
The bolometric luminosity of the stars varied with the LSP phase.
We confirmed that all program stars show the bolometric variations associated with the LSP.

Most stars have $\Delta L/L$ lying between 0.03 and 0.07.
The median values for 9 oxygen-rich stars and 9 carbon stars for $\Delta L/L$ are about 0.046 and 0.053, respectively, where the difference is enough smaller than the range of $\Delta L/L$ values.
This result suggests that there is little to no difference in the typical value for $\Delta L/L$ between oxygen-rich stars and carbon stars.

\begin{figure}
\includegraphics[width=0.5\textwidth]{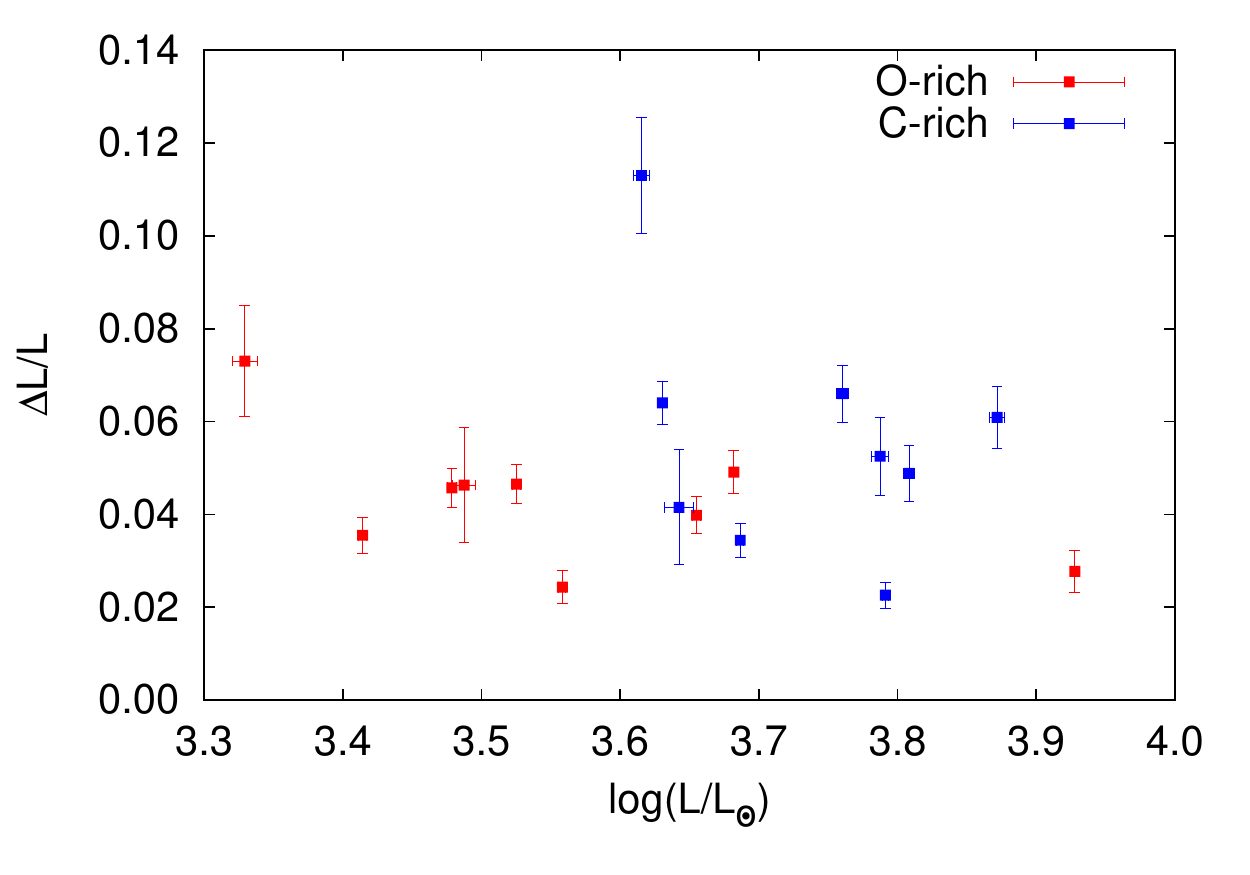}
\caption{Bolometric amplitude plotted against mean luminosity in solar unit.}
\label{L_dL}
\end{figure}

Figure\, \ref{L_dL} shows plots with our program stars for the relation between the mean luminosity ($\log (L /L_{\odot})$) and the bolometric amplitude ($\Delta L /L$). 
The full bolometric amplitudes are twice as large as $\Delta L /L$.
If LSP light variations are caused by an eclipsing binary, the radius of the companion star should be $\sim30\%$ of the radius of the primary (AGB) star in order to reproduce the observed light amplitudes e.g. a system consisting of $\sim130R_{\odot}$ AGB star and $\sim40R_{\odot}$ lower luminosity red giant for the oxygen-rich LSP stars.
Such a binary system would need to have a high mass ratio ($m_{2}/m_{1}\gtrsim$ 0.9) so that the two components were both red giants at the same time.
The fraction of the LSP candidates were approximately 25 - 50 $\%$ against the luminous red giant variables in MACHO and OGLE database. 
If the LSP is caused by eclipse in a binary system with AGB stars, a mass ratio distribution extremely peaked at 1.0 should be required for red giant binaries (e.g. half to all AGB binaries are necessary to be an eclipsing binary with a mass ratio of $\gtrsim$ 0.9 even if binary fraction of AGB stars is 0.5.).
This means that eclipsing binaries cannot explain the LSPs.

\subsection{The radius variations}
\label{sec:radivar}
\begin{figure}
\includegraphics[width=0.5\textwidth]{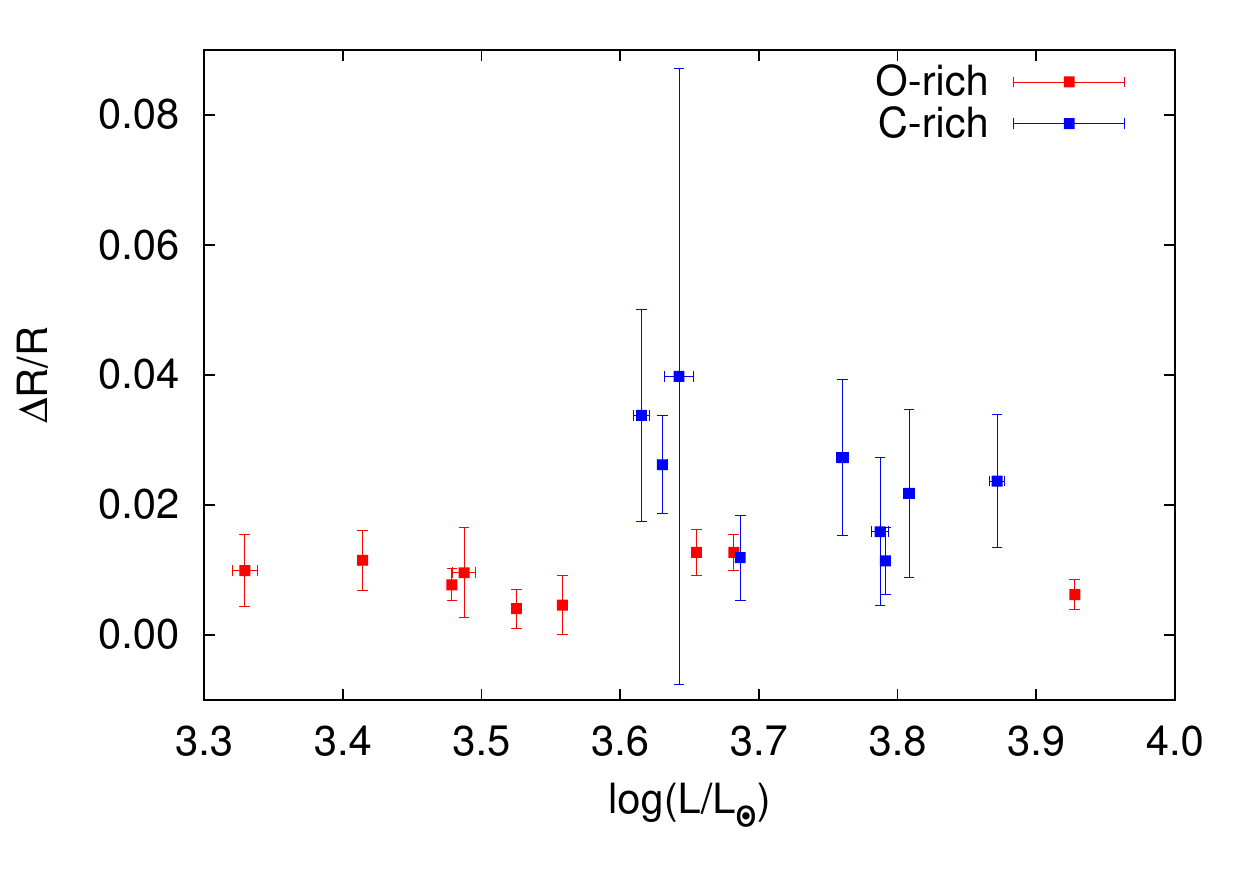}
\caption{Radius amplitude plotted against mean luminosity in solar unit.}
\label{L_dR}
\end{figure}

Figure\, \ref{L_dR} shows plots with our program stars for the relation between the mean luminosity ($\log (L /L_{\odot})$) and the radius amplitude ($\Delta R /R$).
The median values for mean radius for the oxygen-rich stars is about 128$R_{\odot}$. This value is very similar to the typical value for oxygen-rich stars with LSPs in the LMC (cf. \citealt{nic09}).
Most oxygen-rich stars have full radius amplitude (2$\Delta R$) lying between 1$R_{\odot}$ and 4$R_{\odot}$ and the median value is about 2.2$R_{\odot}$.
Those values are slightly smaller than those for oxygen-rich LSP stars in the LMC (cf. \citealt{nic09}).
The carbon stars, on the other hand, have larger radius than the oxygen-rich stars.
Stellar radius for the carbon stars lies in range of 200 - 280$R_{\odot}$ and the typical value is $\sim$250$R_{\odot}$. 
The typical value for full radius amplitude for those stars is about 11.7$R_{\odot}$.

\begin{figure*}
\begin{tabular}{ccc}
\begin{minipage}{0.33\hsize}
\includegraphics[width=1\textwidth]{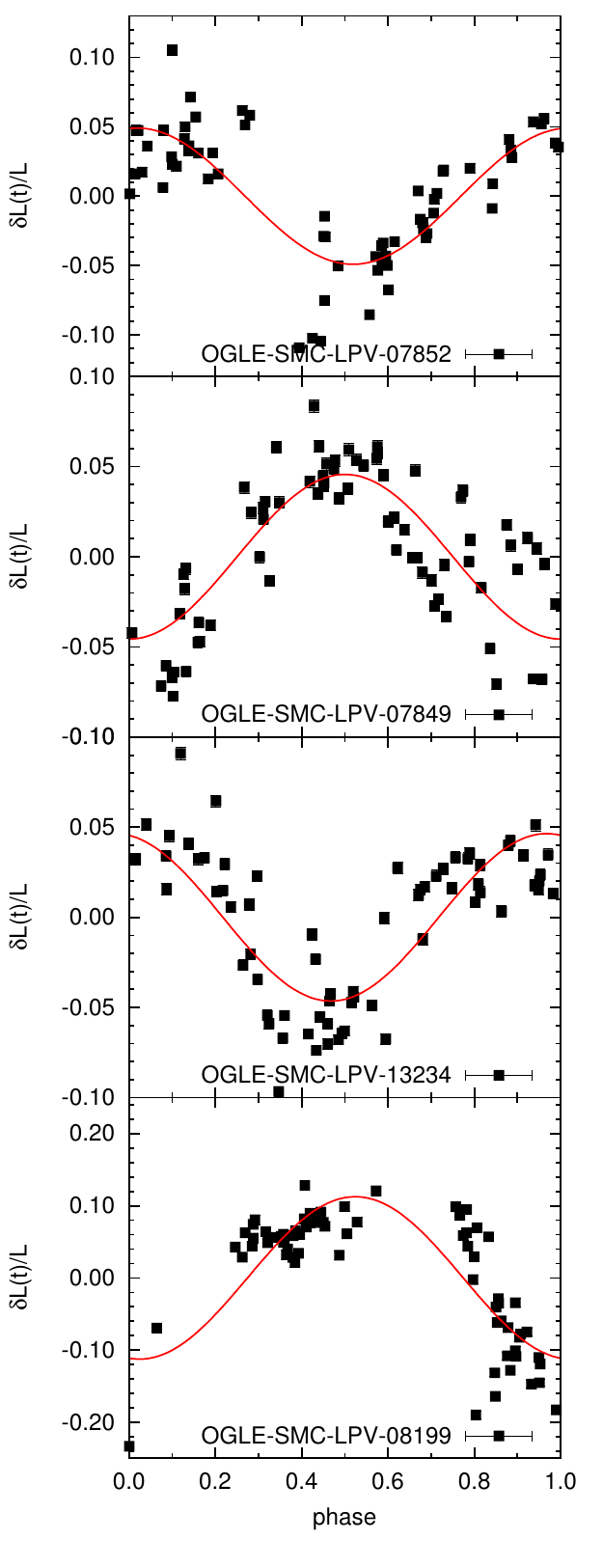}
\end{minipage}
\begin{minipage}{0.33\hsize}
\includegraphics[width=1\textwidth]{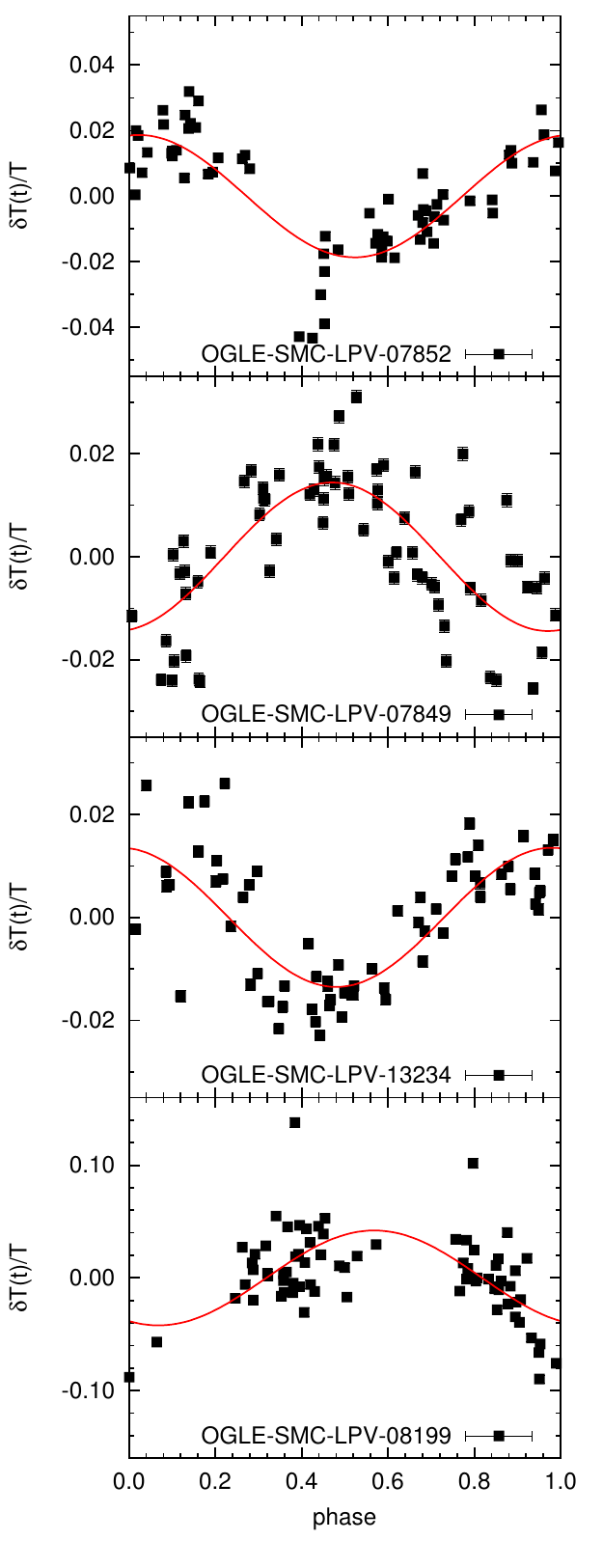}
\end{minipage}
\begin{minipage}{0.33\hsize}
\includegraphics[width=1\textwidth]{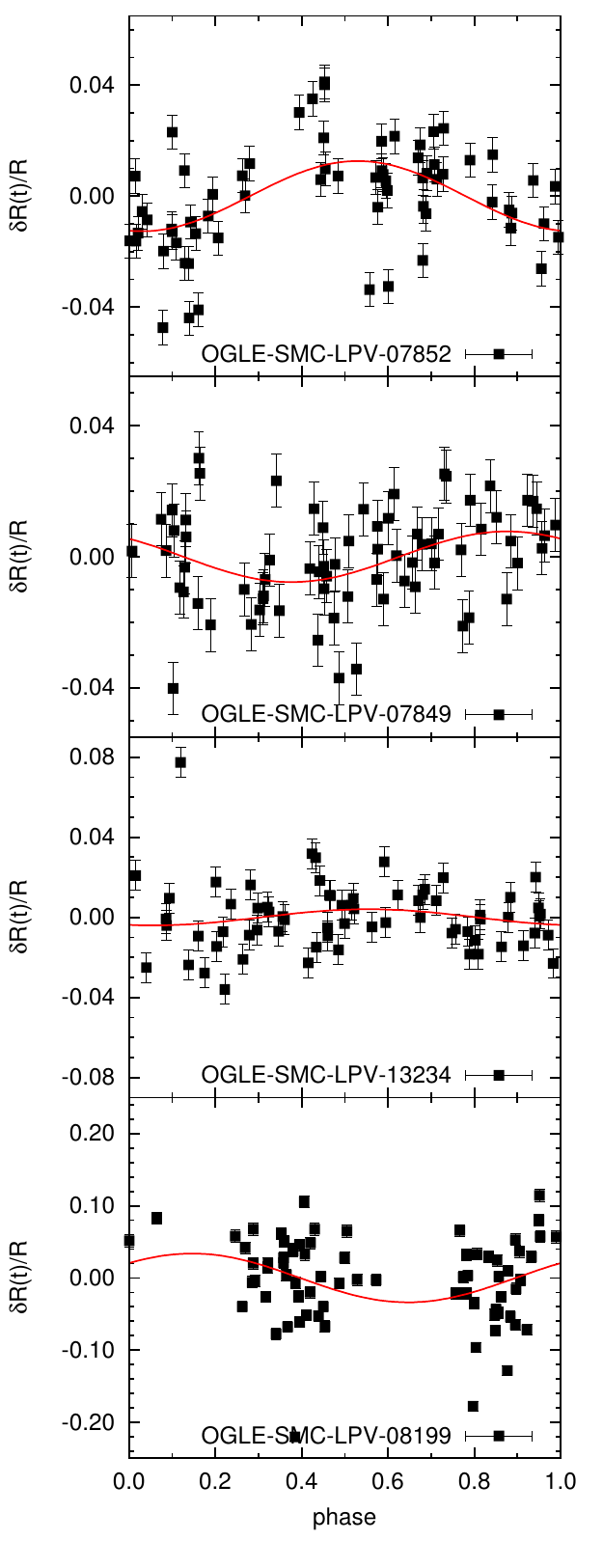}
\end{minipage}
\end{tabular}
\caption{$[Left]$ sample light curves where $\delta L(t)$ corresponds to $L(t)-\left< L \right>$. The red solid lines were made by a least square fit with a single sine curve with $P_{I}$. $[Middle]$ same as the left panels but for the effective temperature.
$[Right]$ same as the left panels but for the radius.}
\label{LTR}
\end{figure*}

\begin{figure*}
\begin{tabular}{cc}
\begin{minipage}{0.5\hsize}
\includegraphics[width=1\textwidth]{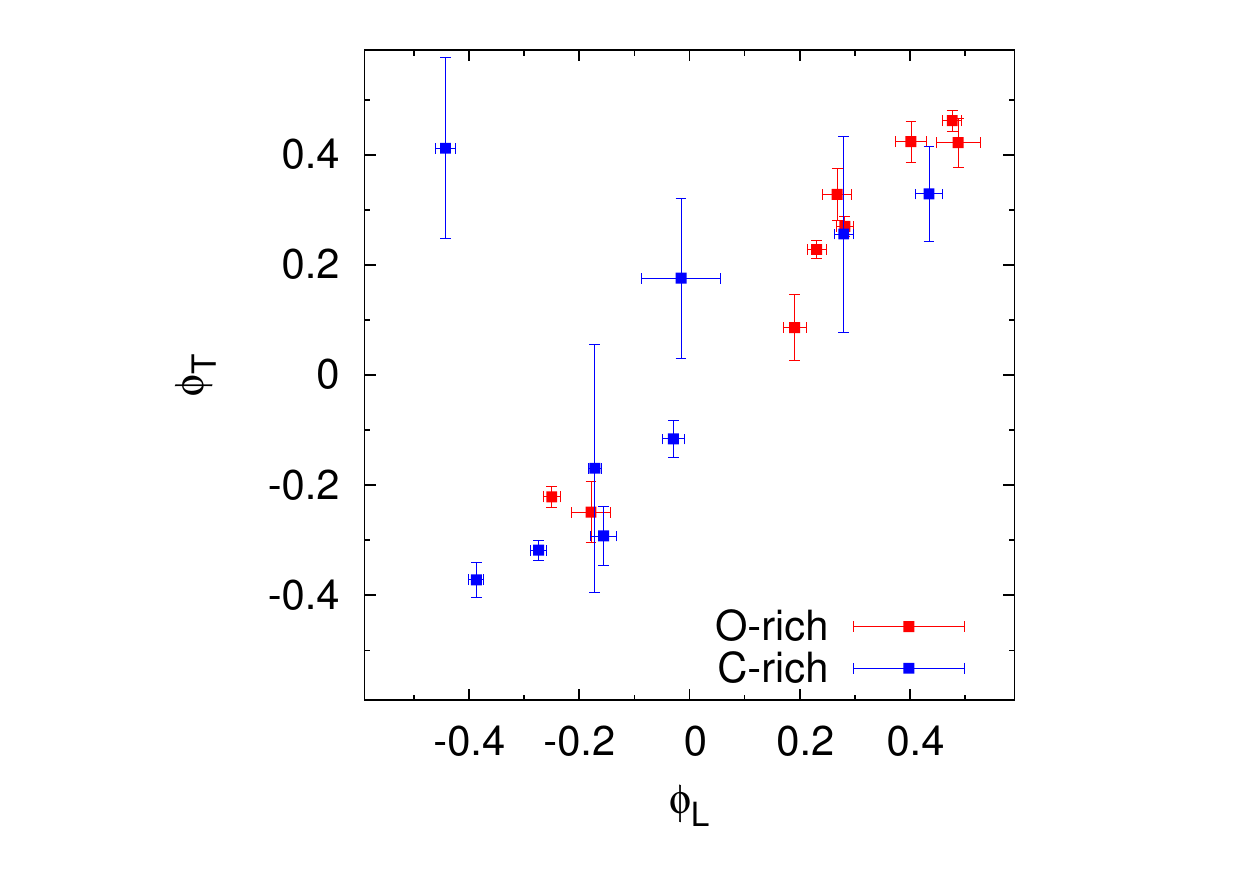}
\end{minipage}
\begin{minipage}{0.5\hsize}
\includegraphics[width=1\textwidth]{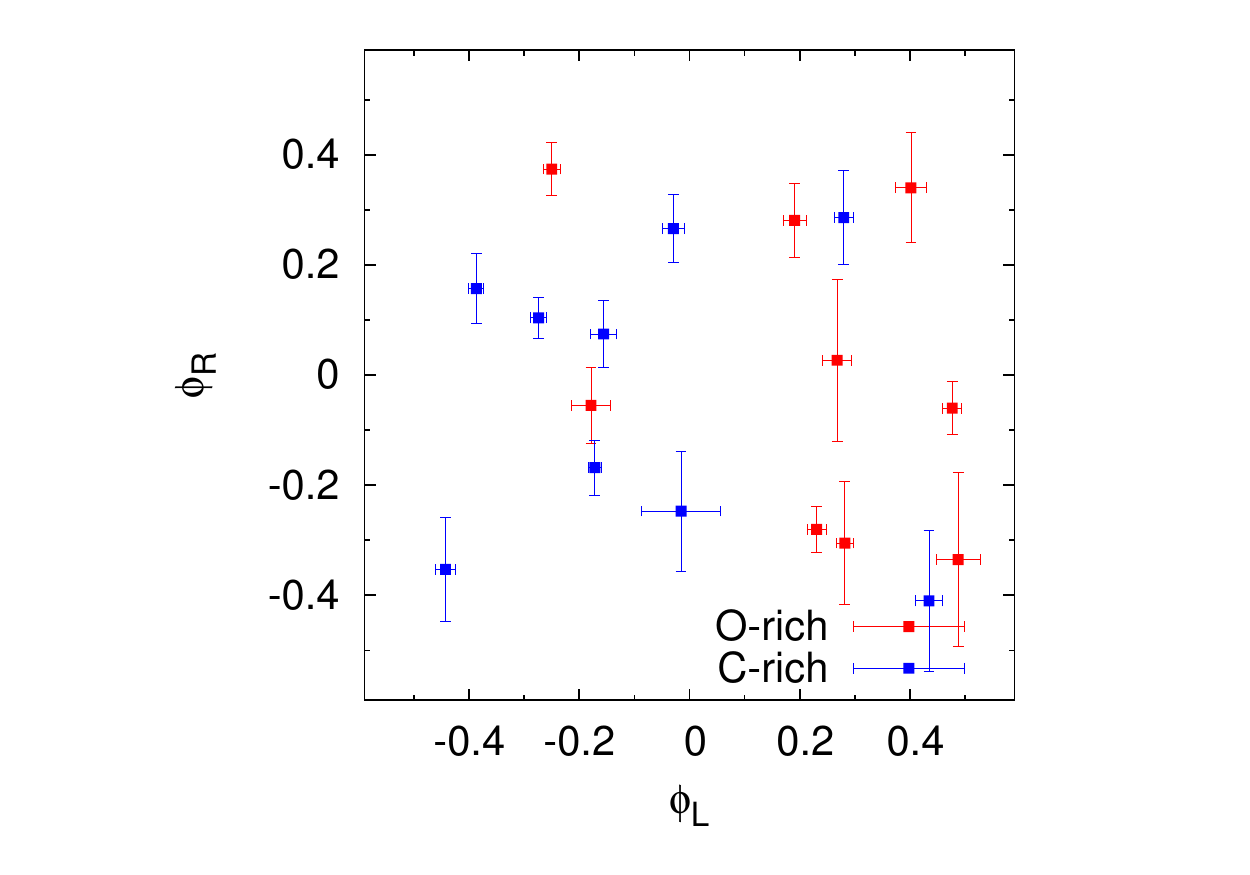}
\end{minipage}
\end{tabular}
\caption{$[Left]$ Comparison of phase for bolometric curves and effective temperature curves. The error bars are derived from the sinusoidal fits. 
$[Right]$ Same as the left panel but for phase of radius curves.}
\label{ph_LRT}
\end{figure*}

We also explored the variation with phase of the radius of our program stars.
Some examples of our sinusoidal fits to the radius data are shown in the right column of Figure\, \ref{LTR}. 
All radius curves shown within this figure are folded at the LSP of the star.  
Similar curves for luminosities and effective temperatures are shown in the left and middle columns, respectively.  
The points of luminosity and temperature lie along the fitted curves while a larger spread around the fitted curve of radius can be seen.
We examined the standard deviation $s_{L}$ between luminosity variations and the fitted curve, 
\begin{eqnarray}
s^{2}_{L}=\frac{\chi^{2}(P_{I})}{N-1}= \frac{1}{N-1} \sum_{i} \left(\frac{\Delta L(t_{i})}{L}-\frac{\Delta L_{P}(t_{i})}{L} \right)^{2}, 
\label{s_L}
\end{eqnarray}
where $\chi$ is defined by Equation \ref{eq_chi}.
The ratio $s_{L}/(\Delta L/L)$ was less than 1.0 in 17 of 18 of our samples. 
The ratio $s_{T}$ and $\Delta T/T$ was less than 1.0 in
5 of 18 of our samples.
However there is no sample star with $s_{R}/(\Delta R/R)<1.0$. 
Therefore radius variations associated with the LSP are obscured compared with variations in $L$ and $T_{\rm eff}$.

The phases of the effective temperature of the stars shown in Figure\, \ref{LTR} seem to correlate with those of the luminosity.
However the phases of the radius are not consistent with the phases of the temperature and luminosity.
The two panels of Figure\, \ref{ph_LRT} show relations between those phases. 
The relative phase $a$ of variations in $L$ and $T_{\rm eff}$ is -0.02±0.02 where $\phi_{T}=\phi_{L}+a$, hence there is no significant phase shift between luminosity and temperature.
On the other hand, the right panel of Figure\, \ref{ph_LRT} suggests no clear consistency in phase between radius and luminosity variations.

\begin{figure}
\includegraphics[width=0.5\textwidth]{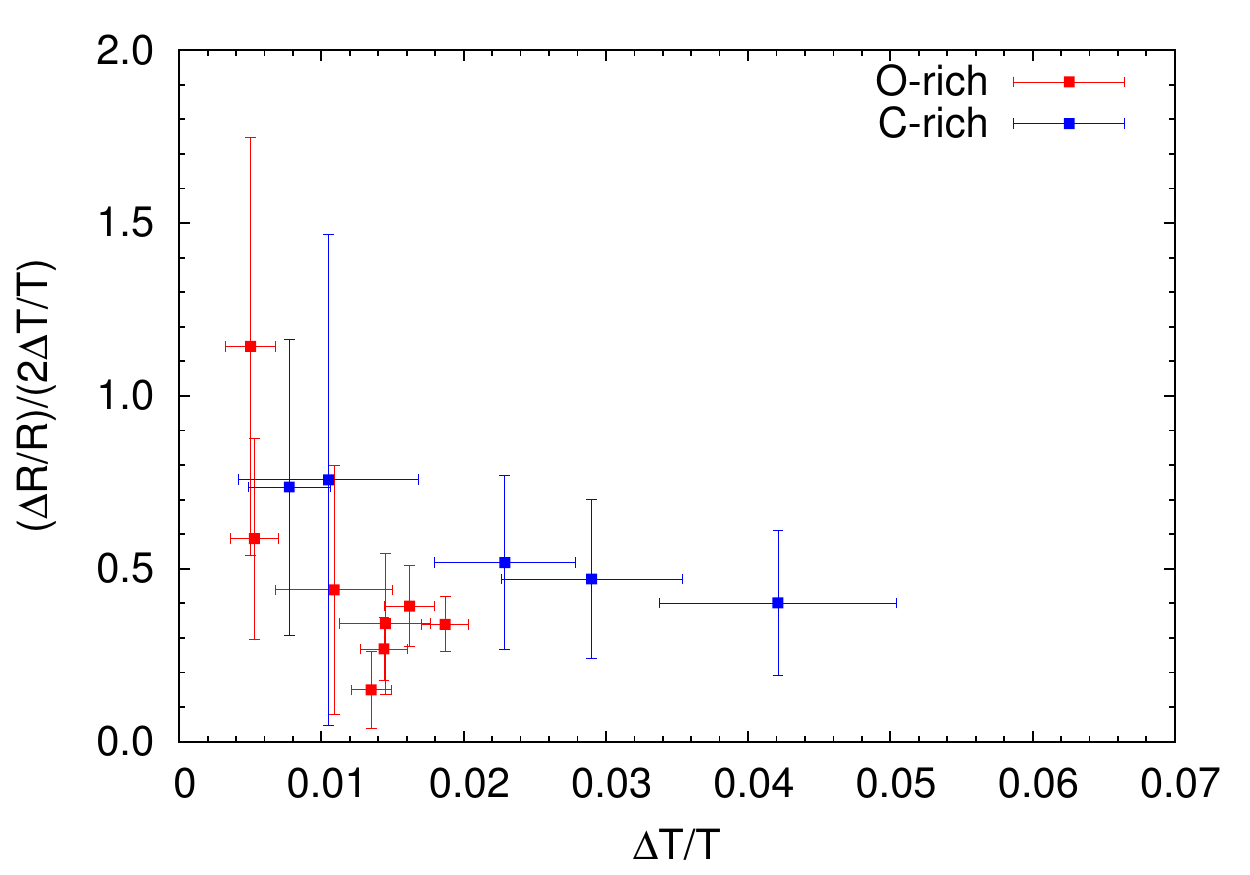}
\caption{Amplitude ratio $(\Delta R/R)/(2\Delta T/T)$ plotted against $2\Delta T/T$ for 13 program stars having $\sigma_{\Delta R}/(\Delta R/R)<0.8$ and $\sigma_{\Delta T}/(\Delta T/T)<0.8$ where $\sigma_{\Delta R}$ and $\sigma_{\Delta T}$ correspond to the error for $\Delta R/R$ and $\Delta T/T$, respectively.}
\label{dT_dRdT}
\end{figure}

We now examine the separate contributions of variations in $T_{\rm eff}$ and $R$
to the observed variation in $L$.
Considering a first order series for perturbations, the Stefan-Boltzmann law is represented by,
\begin{eqnarray}
\frac{\delta L}{L}&=&4\frac{\delta T_{\rm eff}}{T_{\rm eff}}+2\frac{\delta R}{R}\\
&=&4\frac{\delta T_{\rm eff}}{T_{\rm eff}}\left(1+\frac{\delta R/R}{2\delta T_{\rm eff}/T_{\rm eff}} \right).
\label{DLDT}
\end{eqnarray}
Now we compare the amplitude $\Delta R/R$ with $\Delta T/T$ and examine which effect contributes mostly to the luminosity variations.
Figure\, \ref{dT_dRdT} shows the ratio $(\Delta R/R)/(2\Delta T/T)$ against $\Delta T/T$.
Most stars have the ratio with $(\Delta R/R)/(2\Delta T/T) < 1$.
The median value for those ratios for 13 program stars was about 0.44.
Note that only one star, OGLE-SMC-LPV-07858, has $(\Delta R/R)/(2\Delta T/T)$ larger than 1, however, we found no evidence of periodicity for either $R$ or $T_{\rm eff}$.

\begin{figure}
\includegraphics[width=0.5\textwidth]{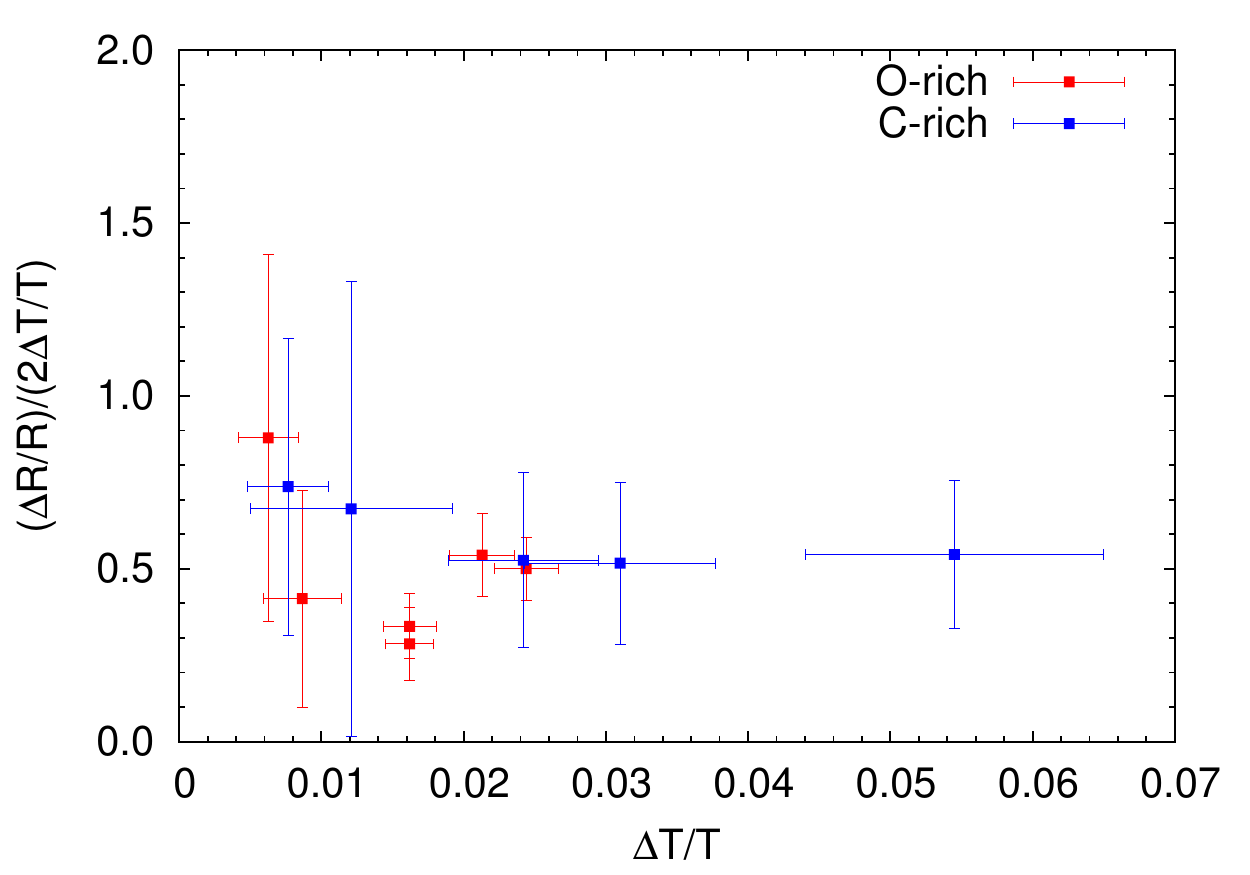}
\caption{Same as Figure\, \ref{dT_dRdT}, but obtained by using the colour - $T_{\rm eff}$ relations of \citet{kuc05} and \citet{mar03} for the oxygen rich stars and carbon stars, respectively. Only the stars with $\sigma_{\Delta R}/(\Delta R/R)<0.8$ and $\sigma_{\Delta T}/(\Delta T/T)<0.8$ are plotted.}
\label{dT_dRdT_r}
\end{figure}

To confirm the robustness for our result, we also calculated the effective temperatures with other colour - $T_{\rm eff}$ relations and reanalysed the amplitudes of $T_{\rm eff}$ and $R$.
The colour - $T_{\rm eff}$ relations of \citet{kuc05} and \citet{mar03} were used for the oxygen-rich stars and carbon stars, respectively.
As a result, we found that the median values for $T_{\rm eff}$ were 3702 K and 3329 K for the oxygen-rich stars and carbon stars, respectively, while the those for $R$ were 129R$_{\odot}$ and 239R$_{\odot}$ for the oxygen rich stars and carbon stars, respectively. 
This suggests that those values are quite consistent with the results obtained by using the colour - $T_{\rm eff}$ relations of \citet{hou00} and \citet{bes83}.
The $(\Delta R/R)/(2\Delta T/T)$ plotted against $\Delta T/T$ is shown in Figure\, \ref{dT_dRdT_r}.
Though the total number of the sample plotted on the $\Delta T/T$ vs $(\Delta R/R)/(2\Delta T/T)$ diagram decreased from 13 to 11, Figures\, \ref{dT_dRdT} and \ref{dT_dRdT_r} show a similar distribution.
Moreover, the the median value of $(\Delta R/R)/(2\Delta T/T)$ for Figure\, \ref{dT_dRdT_r} was about 0.52, thus those values are consistent in the results obtained by different colour - $T_{\rm eff}$ relations.
Hence the $(\Delta R/R)/(2\Delta T/T)$ values were also consistent between different color - $T_{\rm eff}$ relations. 

In conclusion, the change in effective temperature plays an important role in the luminosity change associated with the LSPs.
The change in radius contributes roughly half as much as the change in $T_{\rm eff}$.

\section{discussion}
\label{sec:discs}
\subsection{Possible explanation for the LSP}
\subsubsection{Eclipsing binary}
At first, we argue the inconsistency of the eclipsing binary hypothesis.
Let us consider an eclipsing binary system consisting of a red giant and its companion, which both have no star spot on their surface. 
If the brightness of the companion star can be ignored relative to that of the primary star (red giant), the radius ratio of the red giant to the companion star would determine the depth of the dimming of light due to eclipse.
Moreover, the depth would not vary with waveband.
On the other hand, if the companion's brightness cannot be ignored relative to the red giant's brightness, the contamination by the companion's light would be non-negligible  in the total brightness of the binary system.
The light amplitude variations with wavelength found in the LSP stars can be interpreted as a result that the companion is moderately bright and the effective temperatures differ between the red giant and the companion star.

An eclipsing binary should produce a light minimum at all wavelengths at the same time, thus, at the same phase.
This requires coherence of the light curves in various wavebands.
However, especially in the oxygen-rich stars, we found evidence for the significant phase lag of the light curves between different wavebands.
This result is inconsistent with the explanation of the eclipsing binary.
To explain the LSP variability by eclipsing binary, we would have to consider a more complex binary system such as stars with a star spot and/or binary with a comet-like companion.

\subsubsection{Ellipsoidal binary}
An ellipsoidal binary is one of the possible explanations for the LSPs.
The ellipsoidally distorted shape of a central star gives rise to the light variations associated with the orbital motion. 
The non-spherically-symmetric distribution of the temperature in the photosphere i.e. the gravity darkening effect would also likely cause the periodic variations of the observed temperature.

However, we found that the light amplitudes of the ellipsoidal variability are too small to reproduce the light amplitudes found in the LSPs.
\citet{nie12} found the relations between the orbital elements and the light amplitudes of ellipsoidal binaries with red giants.
We used them to estimate the $I$ amplitudes expected from the LSP stars. 
We assumed a binary system with red giant mass of $1.5M_{\odot}$, companion mass of $0.1M_{\odot}$, and orbital separation $a$=1.4 au, which have been derived from the radial velocity curves of the LSP stars (\citealt{hin02}; \citealt{nic09}). 
The amplitudes vary with the inclination angle of the orbit.
However, when the LSP star has a radius of $R=128R_{\odot}$, the upper limit for $I$ amplitude was only about 0.015 mag.
This is inconsistent with the our result that all our program stars had full $I$ amplitudes of $>$0.08 mag.
Hence ellipsoidal binary hypothesis is inconsistent with the observed light amplitudes in LSP stars.

\subsubsection{Stellar pulsation}
Stellar pulsations on red giants are one of the well-discussed explanations for the LSPs.
As mentioned in Section \ref{sec:intro}, the LSPs are longer than the pulsation periods corresponding to the radial fundamental mode in typical AGB stars (i.e. Mira type stars).
This is one of the reasons of why radial pulsations are ruled out from the explanations for the LSPs.
However, as also shown in Figure\, \ref{L_dR}, we could not find an evidence that the radius amplitudes of our sample stars are zero although the amplitudes were very small.
Therefore, radial pulsation would have room for consideration.

On the other hand, non-radial pulsations are likely to be the explanation for the LSPs.
Periods of some non-radial $g$-modes are longer than period of radial fundamental mode.
Hence, non-radial $g$-modes should be considered for explaining such a long period.
\citet{saio15} suggested that the oscillatory convective modes ($g^{-}$-modes) can be driven in luminous AGB stars.
They also found that the theoretical period-luminosity relations obtained by the dipole oscillatory convective modes are roughly consistent with the sequence D in the LMC.

There is another reason to support non-radial pulsations.
In general, in the dipole mode oscillations described with spherical harmonics, the area of the apparent disc of the star would be maximized twice in one pulsation cycle due to symmetry of a change in a stellar shape.
It means that the period corresponding to variations in $R$ is a half as much as that of $T_{\rm eff}$.
If the contribution of a change in temperature is much larger than that in radius, the period corresponding to luminosity variations would be same as that of the temperature variations.
As mentioned in Sec \ref{sec:radivar}, change in $R$ contribute to change in luminosity about half as much as change in $T_{\rm eff}$. 
It implies that the change of the effective temperature mainly produces the bolometric variations.
The absence of the correlation in relative phase between radius and luminosity (see Figure\, \ref{LTR}) can be interpreted as a result of a fit of $R$ variations with the period corresponding to $L$ variations.
Hence, the observation is consistent with properties of the dipole mode oscillations where the temperature change is main source of the bolometric change.

Also if the LSP variability is caused by dipole mode oscillations of the stars, the observational light amplitudes should vary with the inclination angle of the pulsation axis, though the amplitudes also depend on the pulsation amplitude for the temperature.

In the next section, we investigate whether the dipole mode pulsations in a red giant can reproduce the LSP variability obtained from OGLE and IRSF/SIRIUS survey.
To compare with the observation, we calculate numerical models to obtain the light amplitudes expected from dipole mode oscillations.

\subsection{Comparison of models with observations}
\begin{figure*}
\begin{tabular}{cc}
\begin{minipage}{0.5\hsize}
\includegraphics[width=1\textwidth]{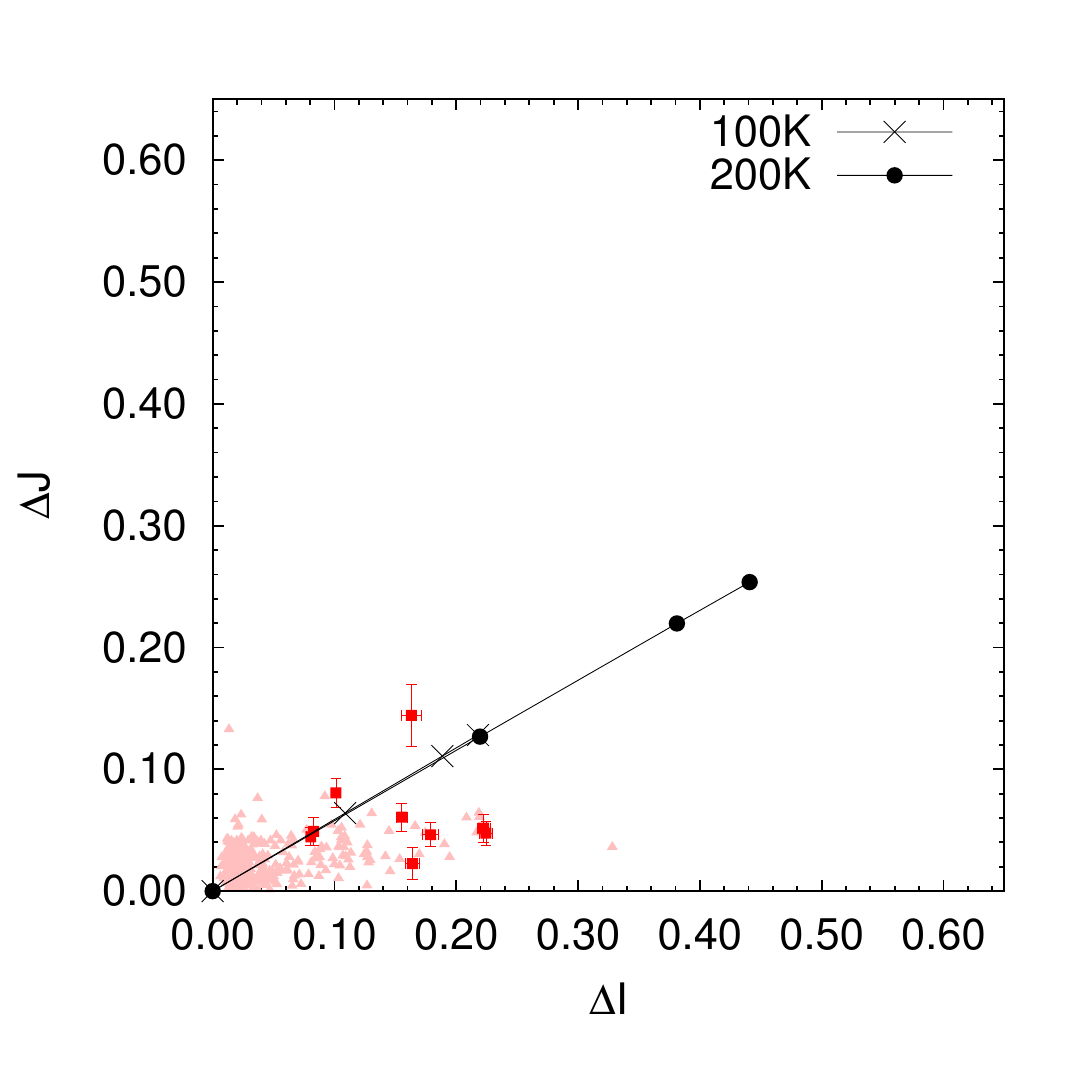}
\end{minipage}
\begin{minipage}{0.5\hsize}
\includegraphics[width=1\textwidth]{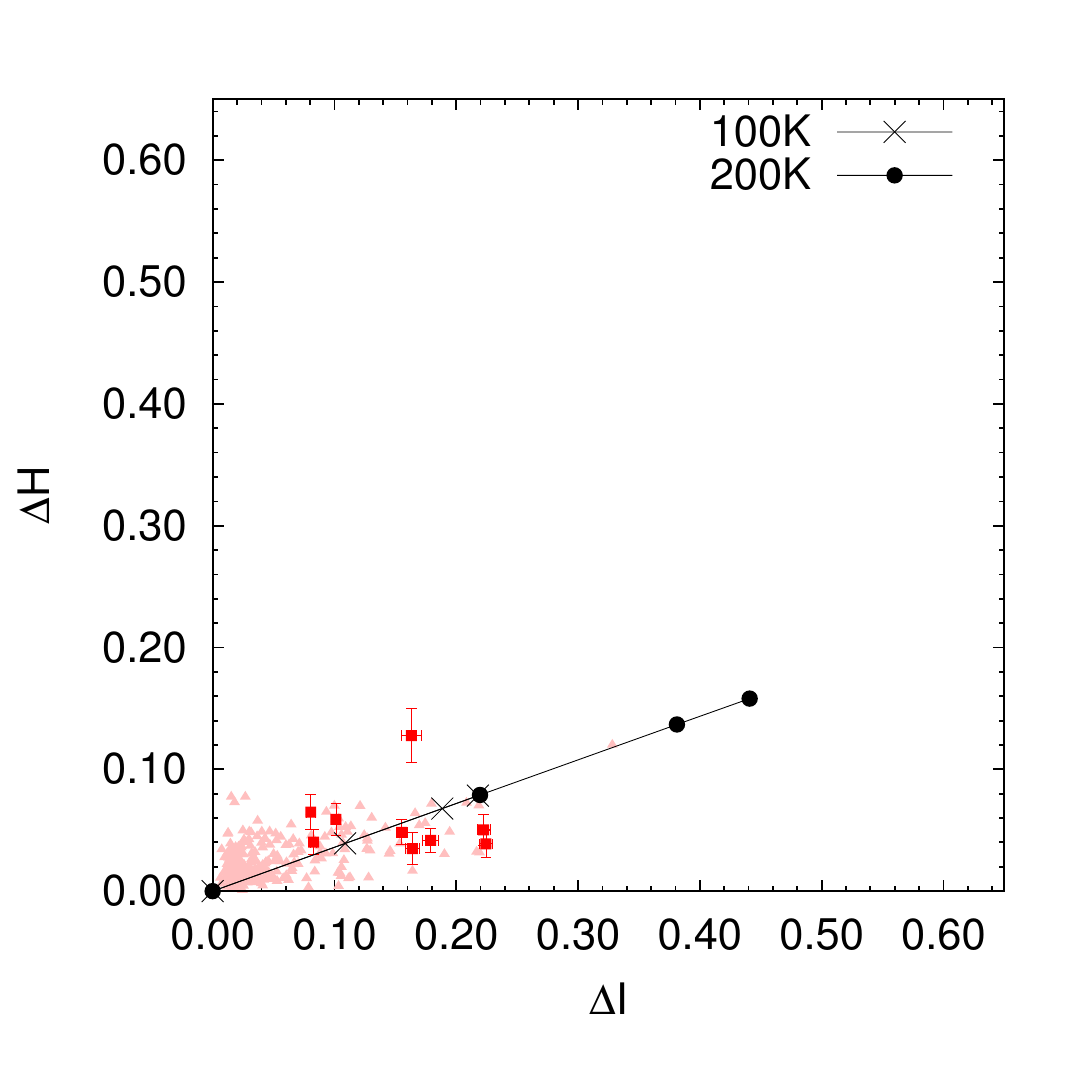}
\end{minipage}\\
\begin{minipage}{0.5\hsize}
\includegraphics[width=1\textwidth]{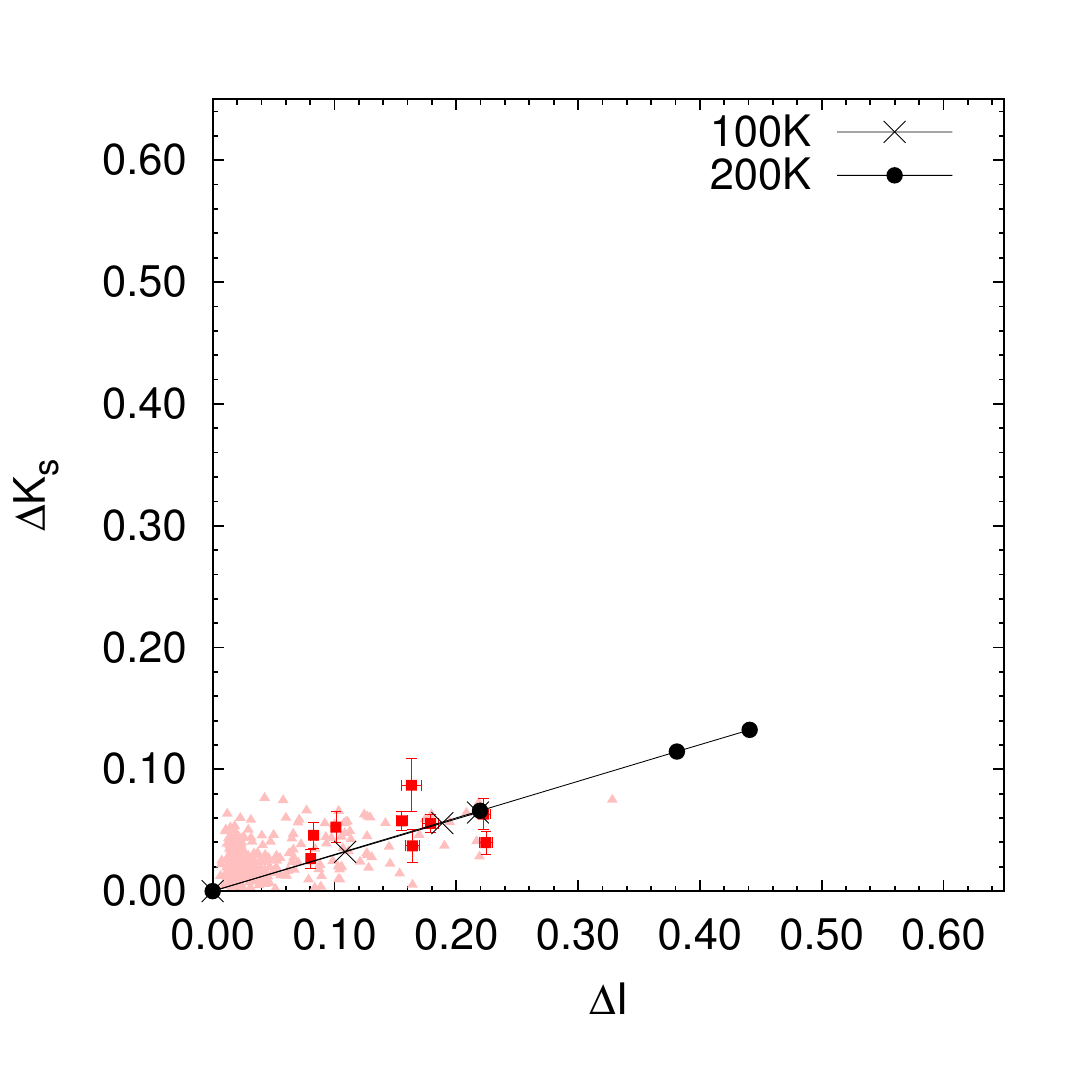}
\end{minipage}
\begin{minipage}{0.5\hsize}
\includegraphics[width=1\textwidth]{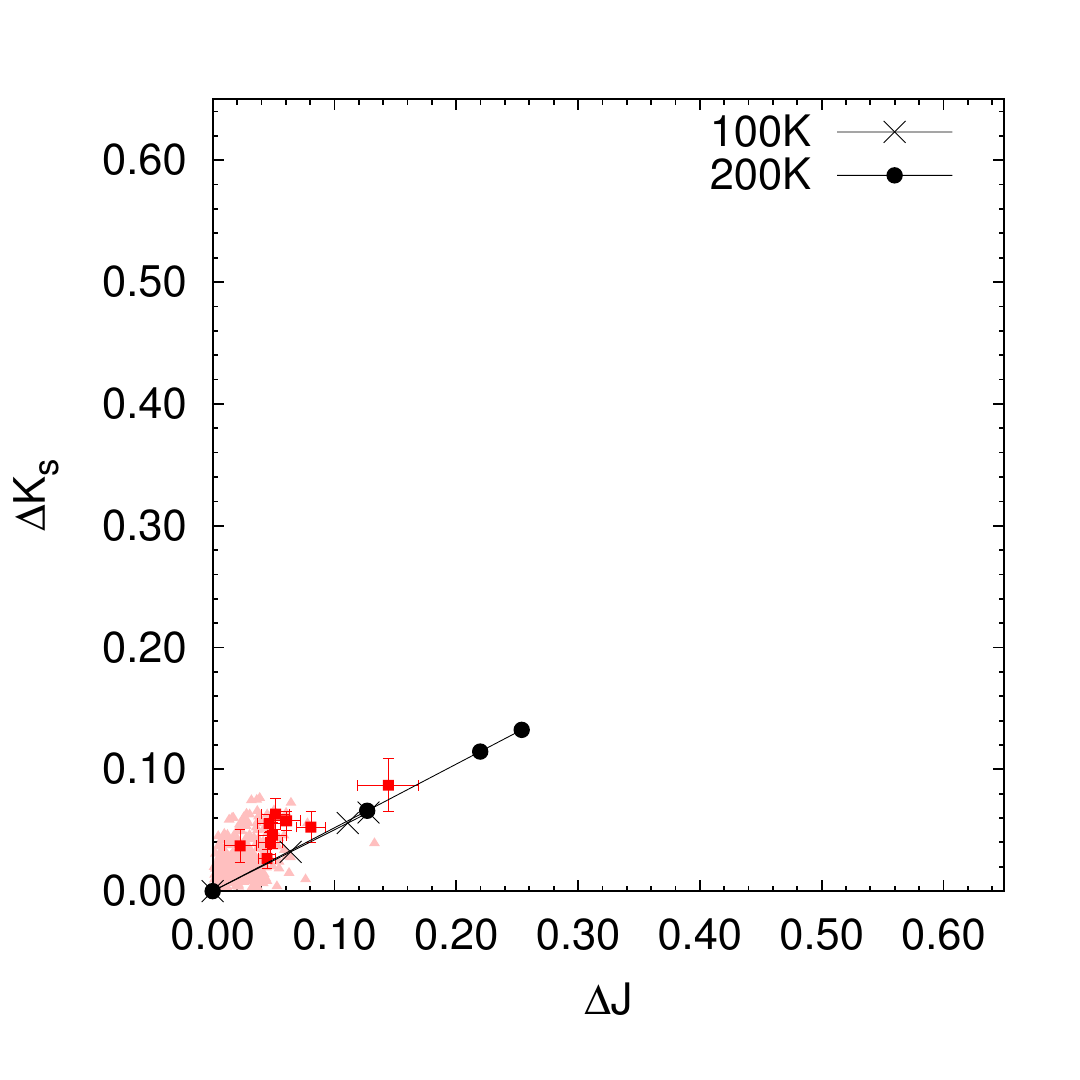}
\end{minipage}
\end{tabular}
\caption{Comparison of models against observation for the full light amplitudes. 
The red filled squares with error bars correspond to 9 oxygen-rich stars. 
The pink-filled triangles show the entire samples which were rejected by our selection criteria for the period analysis (see Sec \ref{select}). 
All light amplitudes are obtained by a fit with a single sine curve with the $P_{I}$.
The black solid lines correspond to the models without limb darkening. 
The black filled circles and crosses show models with a temperature amplitude of 100 K and 200 K, respectively, for inclination angles of 0, 30, 60, and 90 degrees.
}
\label{amp_marcs}
\end{figure*}

\begin{figure*}
\begin{tabular}{cc}
\begin{minipage}{0.5\hsize}
\includegraphics[width=1\textwidth]{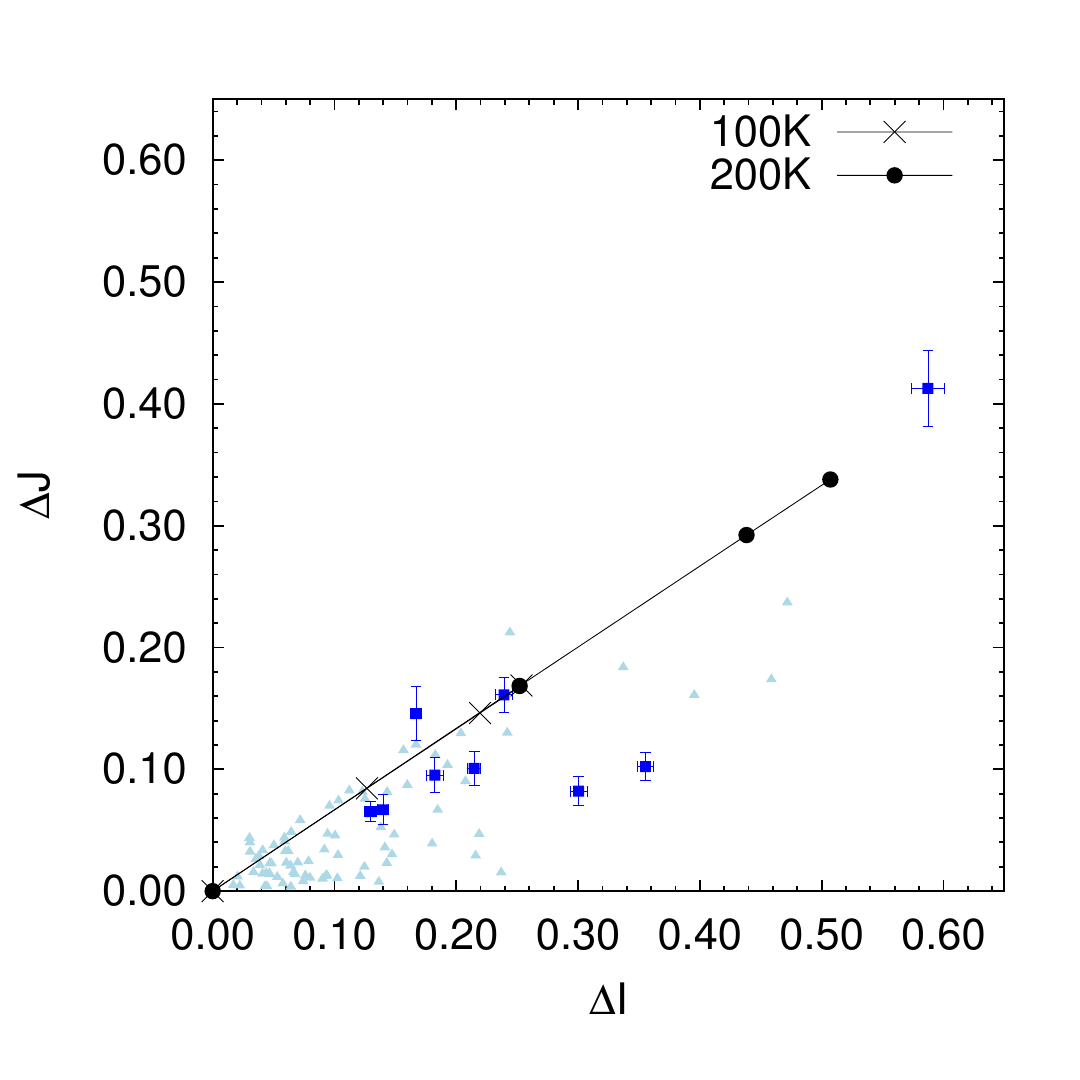}
\end{minipage}
\begin{minipage}{0.5\hsize}
\includegraphics[width=1\textwidth]{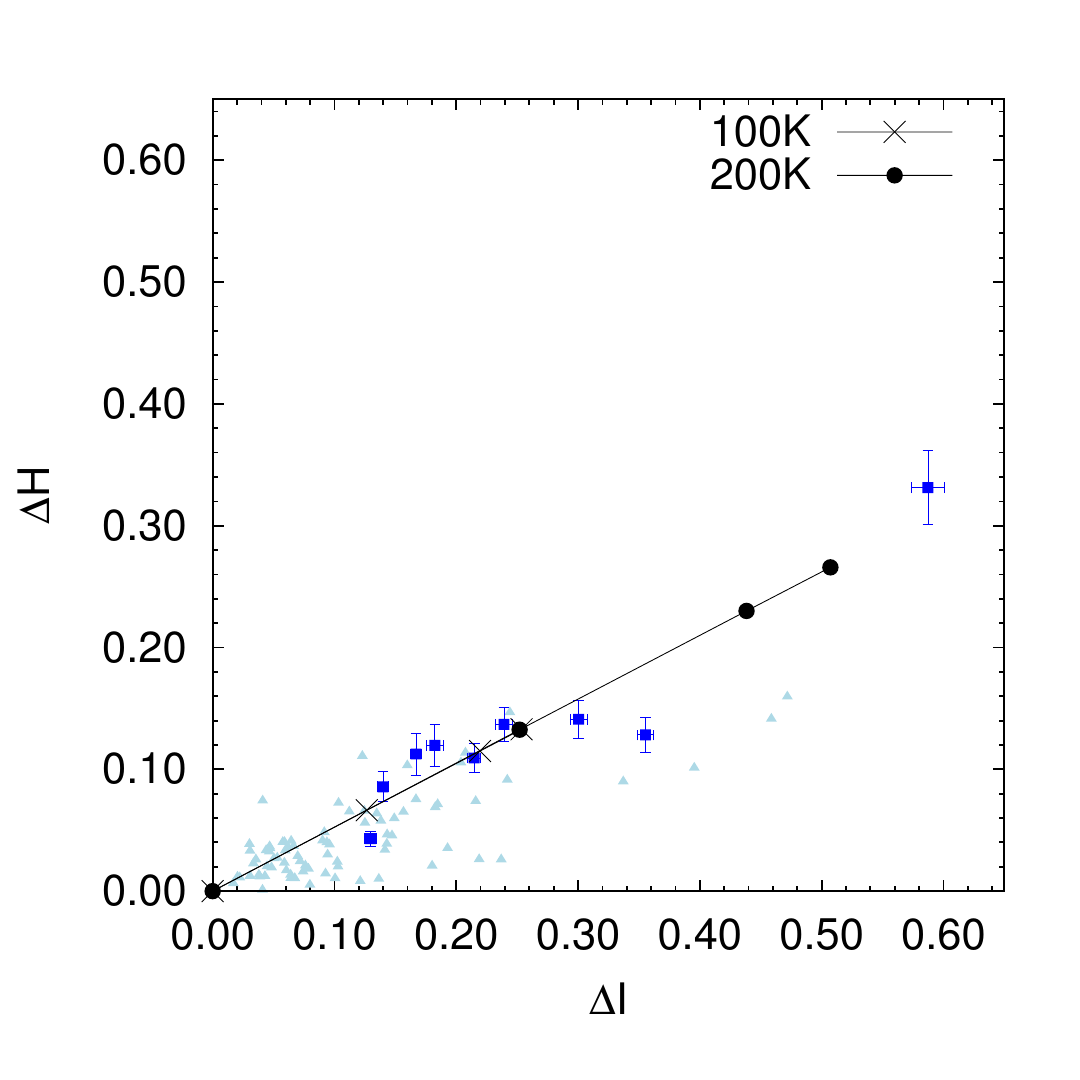}
\end{minipage}\\
\begin{minipage}{0.5\hsize}
\includegraphics[width=1\textwidth]{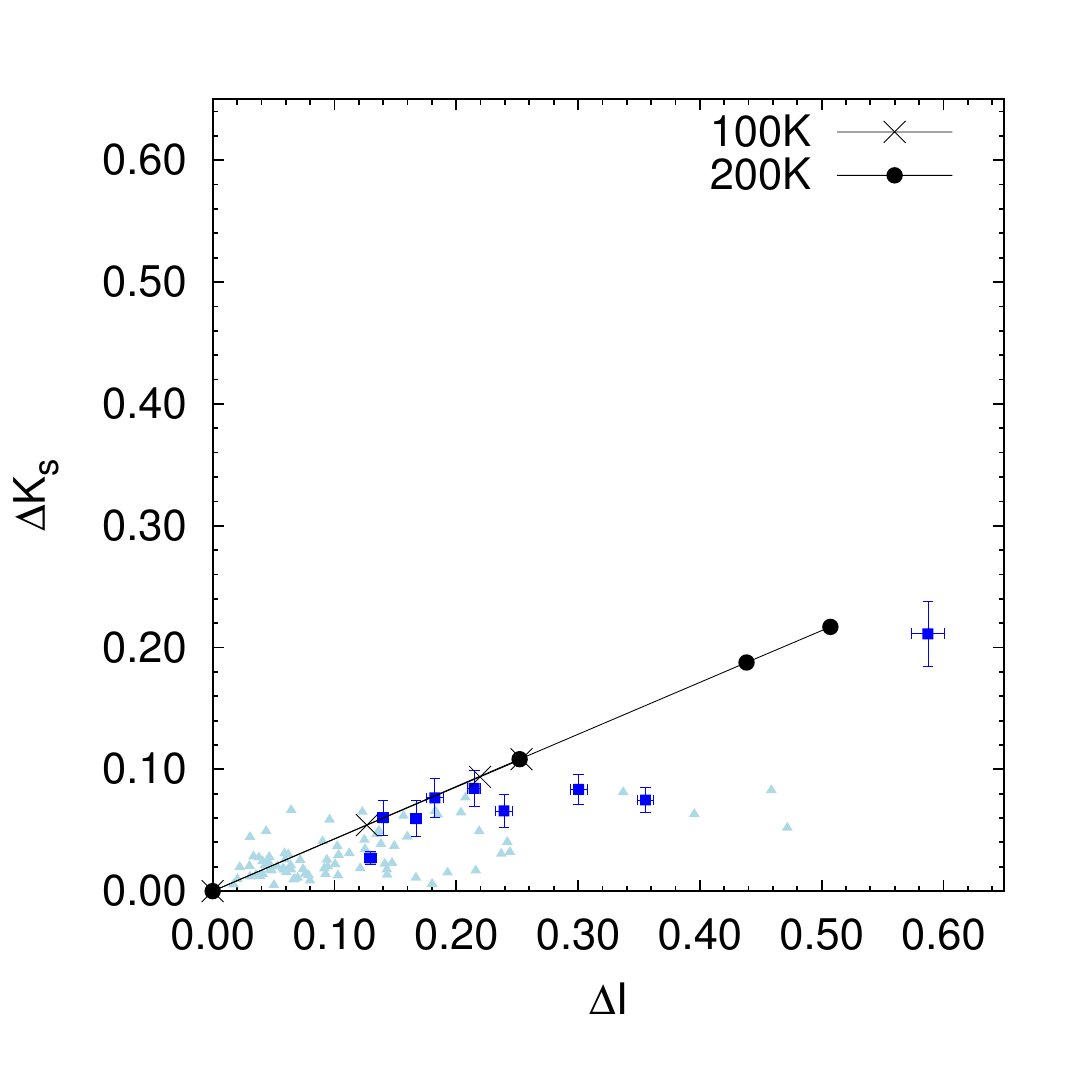}
\end{minipage}
\begin{minipage}{0.5\hsize}
\includegraphics[width=1\textwidth]{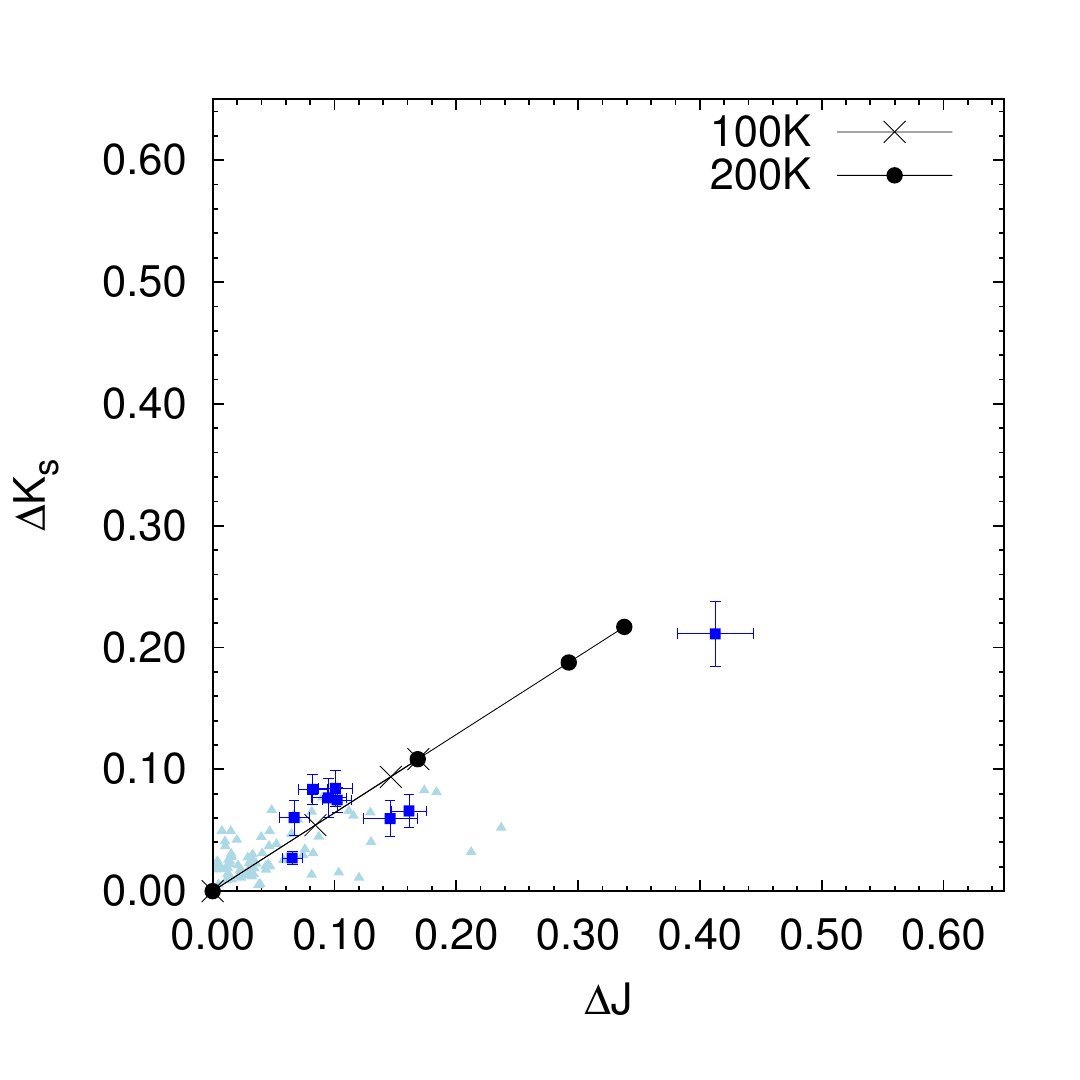}
\end{minipage}
\end{tabular}
\caption{The same as Figure\, \ref{amp_marcs} but the models for carbon stars. 
Again, the blue-filled squares with error bars correspond to 9 carbon stars. 
The light blue filled triangles show the entire samples which were rejected by our selection criteria.}
\label{amp_bb}
\end{figure*}

\begin{table}
\caption{Slopes of the fitted lines and model lines} 
\begin{tabular}{cccc}
\hline
 & slope & observation & model \\
\hline
 & $\Delta J / \Delta I$ & 0.36 $\pm$ 0.08 & 0.59\\ 
O-rich & $\Delta H / \Delta I $ & 0.33 $\pm$ 0.07 & 0.36\\ 
 & $\Delta K_{\rm s} / \Delta I $ & 0.32 $\pm$ 0.04 & 0.30\\ 
 & $ \Delta K_{\rm s} / \Delta J$ & 0.75 $\pm$ 0.08 & 0.50\\ 
\hline
 & $\Delta J / \Delta I$ & 0.56 $\pm$ 0.07 & 0.67\\ 
C-rich & $\Delta H / \Delta I $ & 0.52 $\pm$ 0.03 & 0.52\\ 
 & $\Delta K_{\rm s} / \Delta I $ & 0.32 $\pm$ 0.02 & 0.43\\ 
 & $\Delta K_{\rm s} / \Delta J $ & 0.54 $\pm$ 0.05 & 0.64\\ 
\hline
\end{tabular}
\label{tab03}
\end{table}

Using our models assuming dipole mode oscillations, we calculate the light amplitudes in various waveband and make a comparison with the observations.
At first, we examine models without limb darkening.

Four panels of Figures\,\ref{amp_marcs} shows the plots for a sample of the oxygen-rich stars. 
The model lines are also plotted in the relevant panels.
The maximum amplitudes of each model line correspond to the models for inclination angle of $\alpha$=0 degree.
The temperature amplitudes determine the maximum value of the light amplitudes while the slopes of the model lines are almost the same between the two temperature amplitudes.
Consequently, our models can roughly reproduce the amplitude-amplitude relations of each diagram, though in the ($\Delta I, \Delta J$) diagram the model lines deviate slightly from where many plot points are dense. 

To compare the models with observations quantitatively, we calculated the fitted lines for the plots lying on each panel.
The slopes of the fitted lines and the models with $\Delta T$=100 K are shown in Table \ref{tab03}.
In the plots for ($\Delta I, \Delta H$) and ($\Delta I, \Delta K_{\rm s}$) diagrams, the models can reproduce the slopes of the fitted lines.
The slope for the fitted line for the ($\Delta I, \Delta J$) was 0.36$\pm$0.08, which was gentler than the model slopes by a factor of $\sim$1.6.
For ($\Delta J, \Delta K_{\rm s}$), the slope was 0.75$\pm$0.08 but the model lines were too gentle even if considered 3$\sigma$ error.



These results imply that, in the oxygen-rich stars, the light variations expected from the dipole mode oscillations can reproduce the observed amplitudes for the $I$, $H$, and $K_{\rm s}$ bands, while in the $J$ band, the differences in amplitude between the model and the observation are slightly larger.
A possible interpretation for this discrepancy is the effect of absorption within the wavelength range of the $J$ band by molecules in the stellar atmospheres.
As mentioned in Sec \ref{phaseshift}, the phase variations with wavelength found in the oxygen-rich stars would be likely to be the evidence of the strong molecular absorption at the stellar atmosphere.
Long-term spectroscopic observation is necessary for further investigations.

Four panels of Figures\,\ref{amp_bb} shows the same as Figures\,\ref{amp_marcs} but for the carbon stars.
In contrast with the oxygen-rich stars, we can see apparent positive correlations between the $I$ amplitude and each NIR amplitude.

The model lines were also plotted on the relevant panels.
Note again that the limb darkening effect isn't considered in these models. 
As same as the oxygen-rich stars, the model lines can roughly reproduce the feature of the amplitude-amplitude relations consisting of many plots of the observation.

The slopes of both the model lines and fitted lines of the observations are also shown in Table \ref{tab03}.
The model slopes in the ($\Delta I, \Delta H$) diagram were consistent with the fitted line of the observation.
In the ($\Delta I, \Delta J$) and ($\Delta J, \Delta K_{\rm s}$), the slopes of the model lines and the fitted lines of the observations were in agreement within the margin of 2$\sigma$ error.
However, the model lines could reproduce the observations qualitatively. 
Indeed, those model lines pass through the region where many plots lie on.
For the explanations of the gap between models and observations, more realistic models for the atmospheres of carbon stars are necessary for future study.

\begin{figure*}
\begin{tabular}{cc}
\begin{minipage}{0.5\hsize}
\includegraphics[width=1\textwidth]{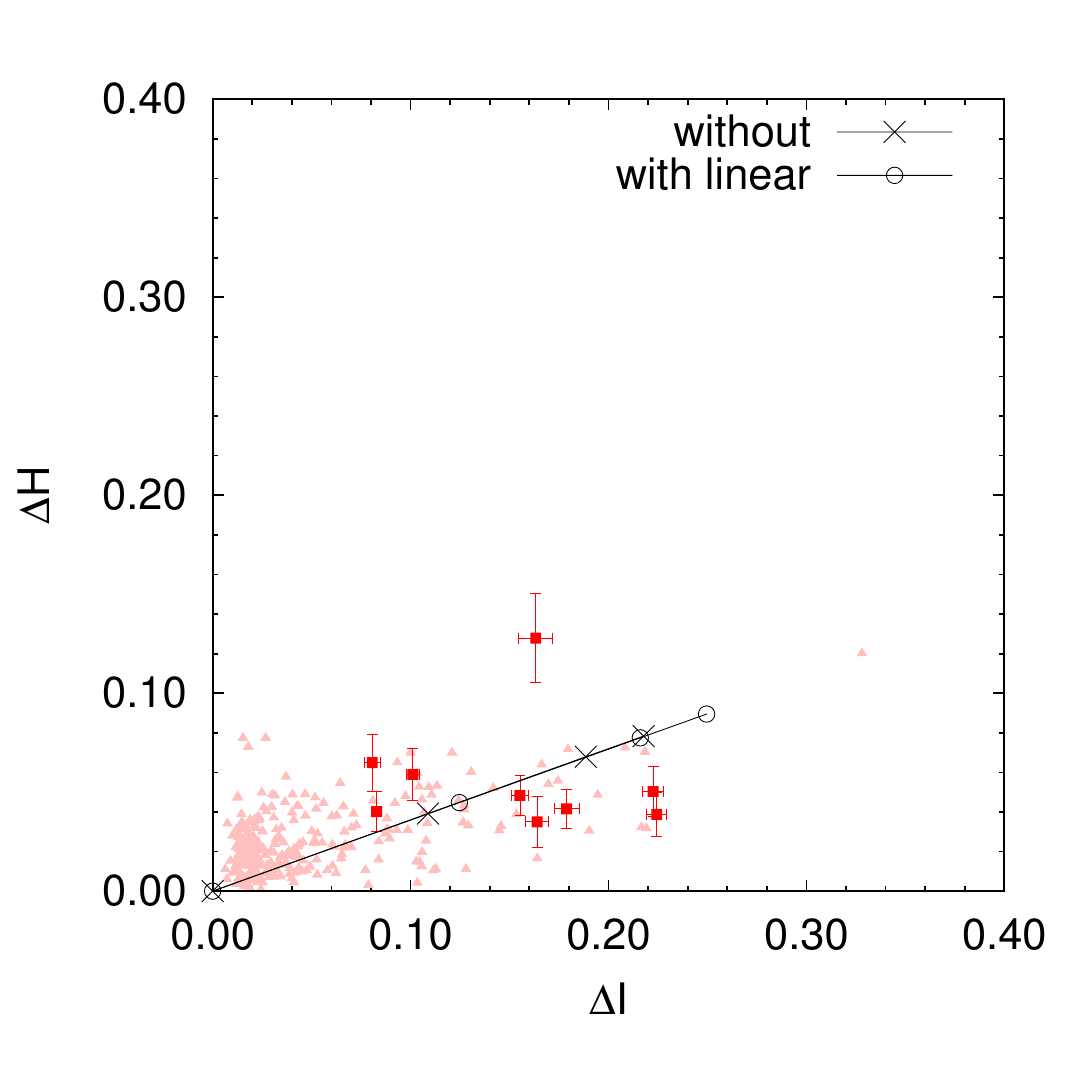}
\end{minipage}
\begin{minipage}{0.5\hsize}
\includegraphics[width=1\textwidth]{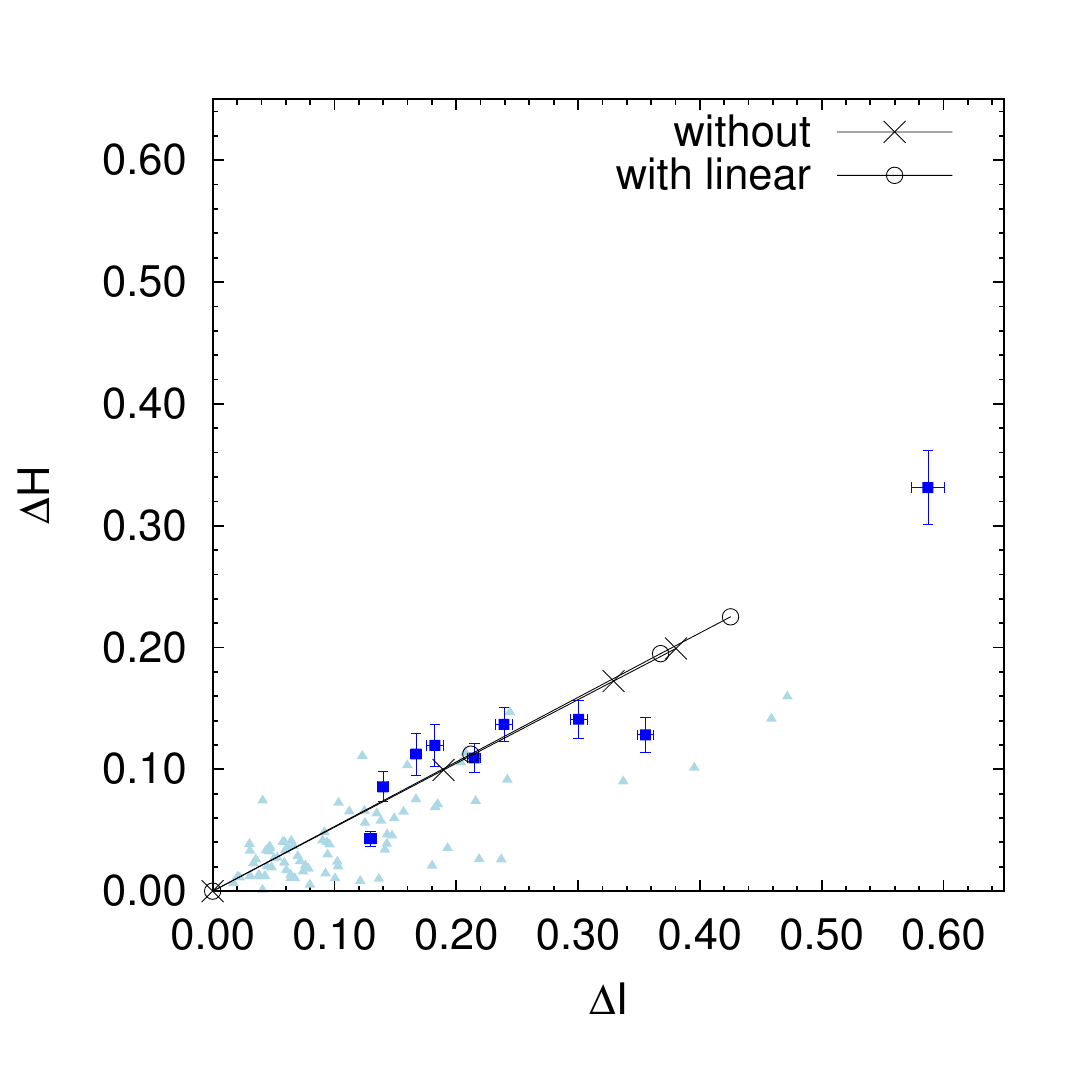}
\end{minipage}
\end{tabular}
\caption{Comparison of the slopes of the model lines obtained from the models with/without limb darkening. The linear limb darkening coefficients obtained by \citet{nei13} were examined for the models. The left panel shows the models for oxygen-rich star with $\Delta T$=100 K. It is same as the left bottom panel of Figure\,\ref{amp_marcs}. The right panel show the models for carbon star with $\Delta T$=150 K. The plots are same as the left bottom panel of Figure\,\ref{amp_bb}.}
\label{comp_amp}
\end{figure*}

We also investigated the effect which the limb darkening gives the model lines.
The two panels of Figure\,\ref{comp_amp} show examples of the amplitude-amplitude relations derived from the models with/without limb darkening. 
It is clear that there is a difference in maximum amplitude between the models with/without limb darkening. 
However, the slopes of those model lines were almost the same. 
We also obtained the same results in the other amplitude-amplitude relations as shown in Figures\,\ref{amp_marcs} and \ref{amp_bb}. 
These results suggest that the models for both oxygen-rich star and carbon star can roughly reproduce the light amplitudes of the LSPs and this conclusion is robust regardless of the limb darkening effect.

As shown in Figure\,\ref{comp_amp}, the maximum values for light amplitudes are altered by the limb darkening coefficients.
With limb darkening, the limb of the stellar disc makes a smaller contribution to the total brightness. 
Moreover, at angle $\alpha$=0, the closer to the limb of the stellar disc, the smaller the amplitude of the brightness. 
Consequently, the intensity perturbations of light from close to the centre of the stellar disc, where the largest amplitude of the intensity perturbations occurs, are emphasized relatively for the total light amplitude of the star. 
This is the reason why the maximum amplitudes for the models with limb darkening are larger than these for the models without limb darkening. 
This also implies that the models with $\alpha$=0 and without limb darkening give us the minimum value of the maximum light amplitudes. 
Hence, the comparison of the observed maximum amplitude with the theoretical
amplitude derived from these models would give us the upper limit of
the temperature amplitude for the pulsations.

We estimated the upper limit for the temperature amplitudes for the pulsations by comparing the observations with the models without limb darkening in the $I$ amplitudes.
In the oxygen-rich stars, the maximum value for the observed $I$ amplitudes was about 0.22 mag and it was consistent with the models for $\alpha$=0 with the temperature amplitude of $\Delta T$=100 K (left panel of Figure\,\ref{comp_amp}).
Hence the upper limit of $\Delta T$ for the oxygen-rich stars would likely to be about 100 K.

The observed temperature amplitude of the stars with $I$ amplitude of 0.22 mag, based on spherically symmetric models, was about 60 K - 70 K (see Tables \ref{tab01} and \ref{tab02}).
We can interpret this smaller value for the amplitude as a result of the spherically symmetric distribution of temperature perturbations, in contrast to the non-spherically symmetric distribution in our models with non-radial oscillations.

The maximum value for the observed $I$ amplitudes for the carbon stars was about 0.59 mag, although most carbon stars have $I$ amplitudes less
than 0.36 mag (right panel of Figure\,\ref{comp_amp}). 
This latter value is consistent with the models for $\alpha$=0 with $\Delta T$ = 150 K.
Hence, we suggest that most carbon stars with LSPs would likely have a temperature amplitude $\Delta T$ less than about 150 K though a small population of carbon stars would have $\Delta T$ larger than 200 K.

\begin{figure*}
\begin{tabular}{cc}
\begin{minipage}{0.5\hsize}
\includegraphics[width=1\textwidth]{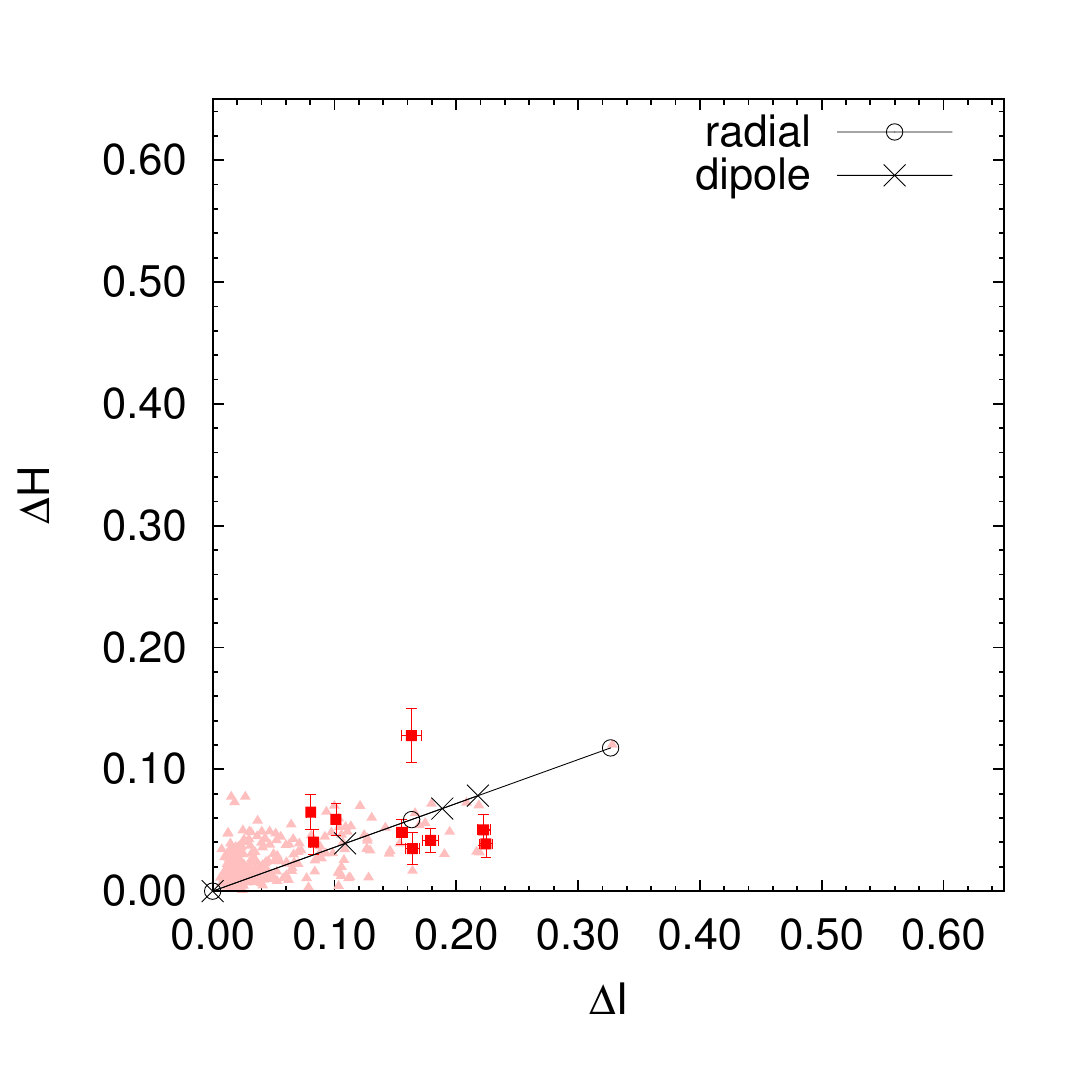}
\end{minipage}
\begin{minipage}{0.5\hsize}
\includegraphics[width=1\textwidth]{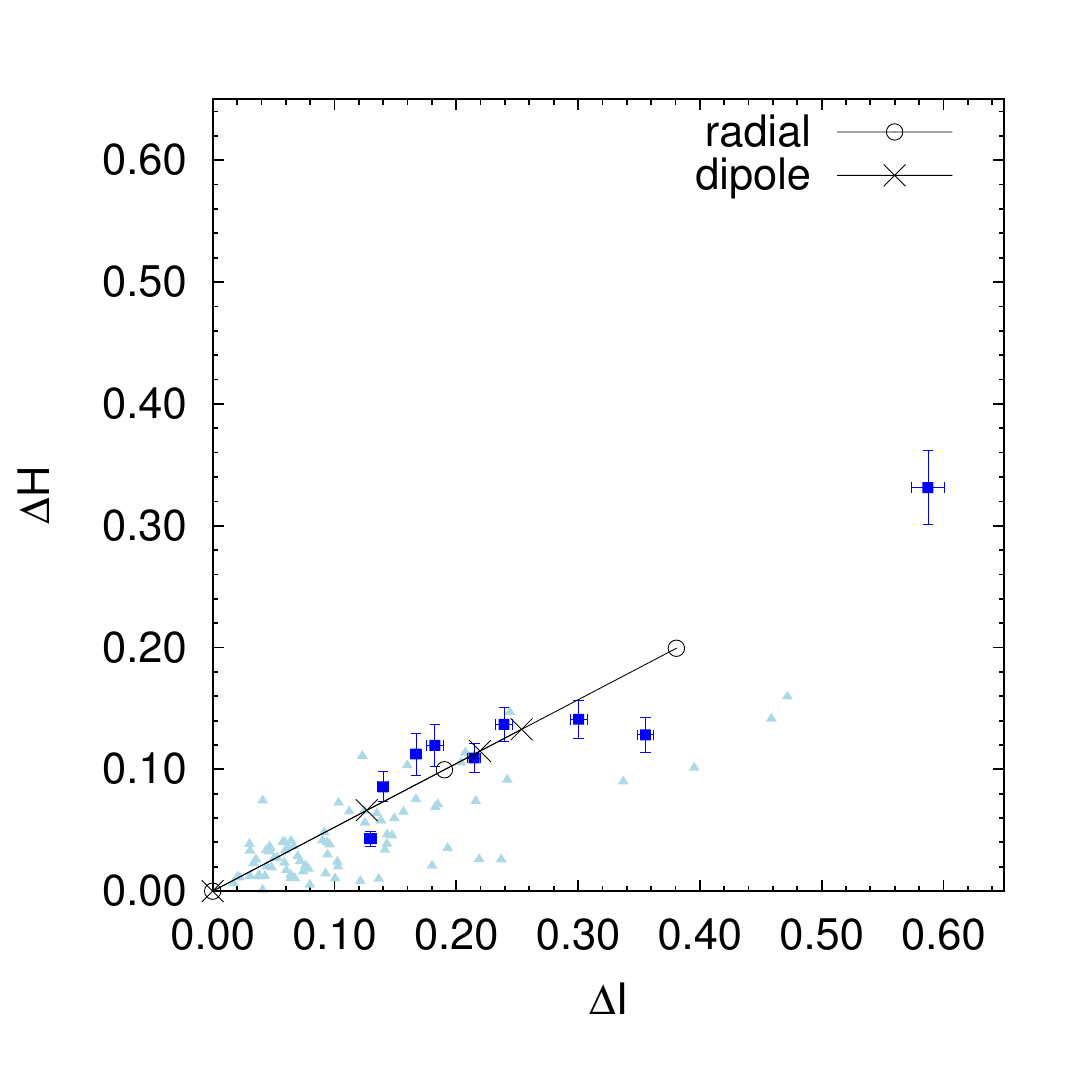}
\end{minipage}
\end{tabular}
\caption{Model lines of radial pulsation for oxygen rich star (left) and carbon star (right). The models for radial pulsations with $\Delta T$=50 and 100 K are plotted. For comparison, the models for dipole mode pulsations with $\Delta T$=100 K which are similar to Figure\,\ref{comp_amp} are also plotted. }
\label{radial}
\end{figure*}

The pulsation period of radial fundamental mode on luminous AGB star has been known to corresponds to the sequence C of the PL relations and it is shorter than the LSPs.
If the LSPs correspond to the radial fundamental mode, the star would be required to have an extremely large radius relative to a typical AGB star.
Nevertheless, we examine light amplitudes of radial pulsations with same stellar parameters assumed for our dipole mode models.

Figure\,\ref{radial} is a comparison between radial pulsation models and dipole mode models. 
The models for radial pulsation are obtained by modification of our dipole mode models.
Also we ignored radius variations for the models hence surface gravity is constant.
Even if non-zero radius amplitude is assumed, the amplitude in a given waveband changes but the slope of the model lines in the amplitude-amplitude diagram would not change when variations in surface gravity are ignorable. 
The slopes of the model lines for radial pulsation are consistent with those of the dipole modes in oxygen-rich and carbon star, respectively. 
The same results were also obtained from other amplitude-amplitude diagrams such as shown in Figure\,\ref{amp_marcs}. 
Those results imply that radial pulsations can not be rejected as the explanation for the LSPs on the basis of relative light variations in optical and near-IR bands.
However, there is still a discrepancy in the length between the LSP and the period of radial fundamental mode of AGB stars. 

\section{conclusion}
We explored properties of the LSP variability and examined whether the dipole mode oscillations of the red giant star can explain the LSPs.
Phase lags between the optical and near-IR light curves of the oxygen-rich stars have been found, which implies that eclipsing binaries are an unlikely explanation for the LSPs.
The evidence for phase lags also implies the possibility of strong molecular absorption in the stellar atmosphere.
A sample of the LSP stars has light amplitude of $\Delta I>$0.08 mag, which would be too large to be explained by ellipsoidal binaries reproducing the typical radial velocity curve of LSP stars.

The effective temperature and luminosity varies during LSP cycles.
The phase difference between variations in $T_{\rm eff}$ and
luminosity is small.
The contribution of the changes in $R$ to bolometric changes is about half as much as changes in $T_{\rm eff}$.
In addition, there is no clear correlation in relative phase between radius and luminosity.
This suggests that variations in $T_{\rm eff}$ make the most important contribution to bolometric change associated with LSPs.
These results are consistent with properties of the dipole mode oscillations of the star but they do not reject radial pulsations from possible explanations for the LSPs.

We have created numerical models for stellar photosphere and computed light amplitudes expected from the dipole mode oscillations.
By comparison with the observations, it has been found that our models can roughly reproduce the amplitude - amplitude relations obtained by the LSP stars.
The diversity in light amplitudes among the LSP stars is likely to be a combination of variable intrinsic amplitude and different inclination angles of the pulsation axis
to the line of sight.
The temperature amplitudes in the photosphere are expected to be mostly less than 100 K and 150 K for the oxygen-rich stars and carbon stars, respectively.
On the other hand, a discrepancy between the gradient of the observed amplitude - amplitude relations and the slopes of the model lines has been found.
In this work, hydrostatic models are assumed for the atmosphere and the spectra although a pulsating star is considered.
Hence, hydrodynamic models for non-radial oscillations in AGB stars might produce better agreement with observations.

Similar to the calculations with dipole mode models, we also have examined photometric amplitude with the radial pulsation models.
The model lines of both pulsation models show quite similar slope on amplitude - amplitude diagrams.
This means that radial and non-radial (dipole mode) pulsations would be consistent with colour changes in LSP stars.
However, there is still an inconsistency in length between the LSP and periods of radial fundamental mode.

\citet{saio15} found that dipole modes corresponding to non-adiabatic $g^{-}$ modes (so called oscillatory convective modes) can roughly reproduce the PL relation of the sequence D.
Therefore our results suggest that the observations can be consistent with stellar pulsations corresponding to oscillatory convective modes.
In order to reproduce the observational results more precisely, it is necessary to construct models taking more complex physics into account, such as an effect of molecular absorption in the atmospheres.

\section*{Acknowledgements}
The authors would like to thank the anonymous referee whose constructive and helpful comments have improved the readability of the paper.
We are  grateful to people involved in the OGLE, $SAGE$ and MCPS projects for making their data so easy to access and use. 
The authors would like to thank the members of IRSF/SIRIUS team for providing their data. 
We are also grateful to the authors of MARCS code. 
We thank Prof. Hedeyuki Saio, Prof. Hiromoto Shibahashi and Dr. Masao Takata for the valuable comments. 
This research is supported in part by the Japan Society for the Promotion of Science through Grant- in-Aid for Scientific Research 26·5091.

\bsp

\label{lastpage}

\appendix
\section{Calculation for light amplitudes expected for the dipole mode oscillations}    
\label{apA}
In order to calculate the light amplitudes expected for dipole mode oscillations of the star, we make very simple models to simulate the light variations in various wavebands. 
We consider linear pulsations on a non-rotating star and assume the spherical harmonics $Y^{m}_{l}(\theta,\varphi)$ for the angular displacements, thus the temperatures of the atmospheres at $(\theta,\varphi)$ are represented by $Y^{m}_{l}(\theta,\varphi)$.

\begin{figure}
\includegraphics[width=0.5\textwidth]{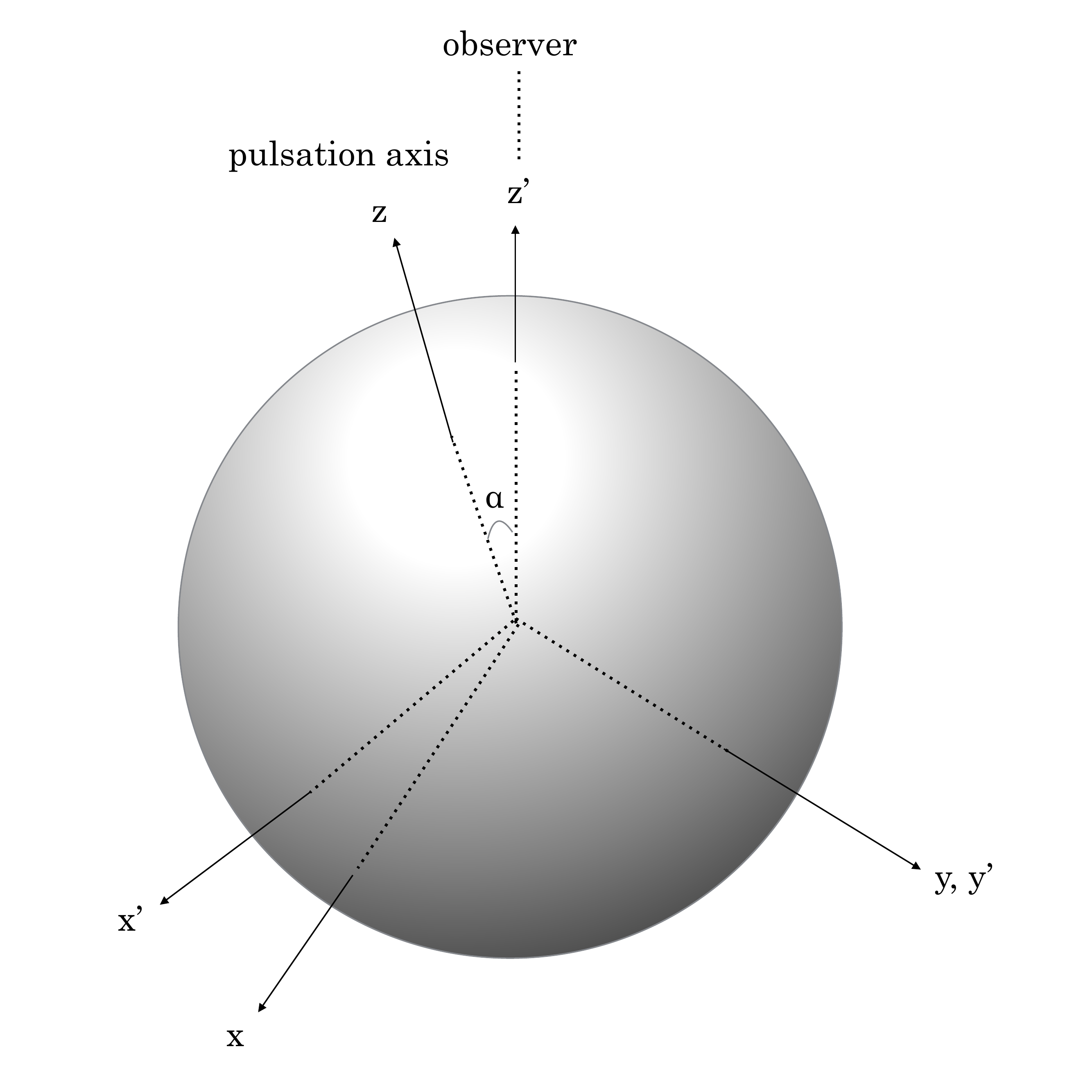}
\caption{An illustration of the coordinate systems for the pulsating star and the observer.}
\label{coord}
\end{figure}

The form of the spherical harmonics is also determined by the angle between the pulsation axis and the line of sight direction from the observer. 
When the pulsation axis corresponding to $\theta=0$ is inclined at angle $\alpha$ to the line of sight direction toward the observer,  the spherical harmonics $Y^{m}_{l}(\theta,\varphi)$ can be represented as,
\begin{eqnarray}
Y^{m}_{l}(\theta, \varphi)=\sum^{l}_{m'=-l} d^{l}_{mm'}(\alpha)Y^{m'}_{l}(\theta ', \varphi ').
\label{SH}
\end{eqnarray}
We set up a new coordinate system $(x',y',z')$ as that the axis with $\theta '$=0 (the $z'$-axis) agrees with the line of sight direction toward the observer and the $y'$-axis agrees with the $y$-axis, i.e. the observer lies on the $z-x$ plane (Figure\, \ref{coord}).

Then the $Y^{0}_{1}(\theta, \varphi)$ can be given by following equation,
\begin{eqnarray}
Y^{0}_{1}(\theta, \varphi)&=&\cos \alpha Y^{0}_{1}(\theta ', \varphi ') \nonumber \\
& &-\frac{1}{\sqrt{2}}  [Y^{1}_{1}(\theta ', \varphi ')-Y^{-1}_{1}(\theta ', \varphi ')] \sin \alpha \nonumber \\
&=&\sqrt{\frac{3}{4 \pi}} (\cos \alpha \cos \theta ' + \sin \alpha \sin \theta ' \cos \varphi ').
\label{y01}
\end{eqnarray}  
We consider $Y^{0}_{1}$ for our simulations because the displacement pattern of $Y^{\pm 1}_{1}(\theta, \varphi)$ can be represented by a geometrical rotation of $Y^{0}_{1}(\theta, \varphi)$.
Then the temperature of the atmospheres at time $t$ can be represented as following equation,
\begin{eqnarray}
T_{p}(\alpha, \theta', \varphi')&=&T_{0}+\delta T(t) Y^{0}_{1}(\theta, \varphi) \nonumber \\
&=&T_{0}+\Delta T (\cos \alpha \cos \theta ' + \sin \alpha \sin \theta ' \cos \varphi ') e^{i\omega t},\nonumber \\
\label{Tp}
\end{eqnarray}
where $T_{0}$ is the temperature at the hydrostatic equilibrium and $\omega$ is the pulsation frequency.

According to the oscillations for dipole modes, the diameter of the star is a constant though the shape of the star becomes a slightly distorted sphere.
For the sake of simplicity, we assume a star keeping a spherically symmetric shape with a constant stellar radius.
Hence the radiative flux $F_{\lambda}$ emitted from the entire stellar disc to the observer for wavelength $\lambda$ can be given by the following equation,
\begin{eqnarray}
F_{\lambda} &=& \frac{1}{d^2} \int_{\rm hemisphere}  I_{\lambda}(\alpha, \theta ',\varphi ') {\bm e_{z'}} \cdot d{\bm s}'\nonumber \\
&=& \frac{R^{2}}{d^2} \int_{0}^{\pi/2} \int_{0}^{2\pi}  I_{\lambda}(\alpha, \theta ',\varphi ') \sin \theta ' \cos \theta ' d\theta ' d\varphi',
\end{eqnarray}
where $I_{\lambda}(\alpha, \theta ',\varphi ')$ is intensity of the brightness on the stellar disc. 
$R$ and $d$ are the stellar radius and the distance from the star to the observer, respectively.
However $R$ and $d$ values are not considered for derivation of the light amplitudes because those are cancelled during the calculation.

The limb darkening effect is also one of the factors that determine the brightness of the star.
For more detail, the limb darkening on a pulsating star with a non-spherical distribution of the temperatures of the atmospheres should be considered. 
Nevertheless, for simplicity, we assumed the limb darkening for single temperature  represented as
\begin{eqnarray}
I_{\lambda}(\alpha, \theta ',\varphi ') = I_{0,\lambda}(T_{p}(\alpha, \theta', \varphi')) D_{T_{0}}(\theta '),
\end{eqnarray}
where $D_{T_{0}}(\theta ')$ is the limb darkening corresponding to temperature of $T_{0}$.
We consider the models for no limb darkening effect (i.e. $D_{T_{0}}(\theta ')=1$) and the models with the linear limb darkening coefficients ($D_{T_{0}}(\theta ')=1-u(1-\cos \theta ')$) of \citet{nei13}.
We also assumed the model intensities of the surface brightness $I_{0,\lambda}$.
For oxygen-rich stars, we assumed the temperature-dependent intensity $I_{0,\lambda}(T)$ given by the spherically-symmetric MARCS code (\citealt{gus08}) models.
On the other hand, we assumed the intensities corresponding to the blackbody for carbon star models.

Finally, the brightness for the waveband is derived by considering the filter function  corresponding to the waveband $f_{\rm band}(\lambda)$,
\begin{eqnarray}
B_{\rm band}=\int f_{\rm band}(\lambda) F_{\lambda} d\lambda.
\end{eqnarray}
The full light amplitudes are derived from the computed brightness in maximum and minimum light.


\begin{thebibliography}{99}
\bibitem[\protect\citeauthoryear{Bessell, Wood, \& Evans}{1983}]{bes83}
Bessell M.~S., Wood P.~R., Evans T.~L., 1983, MNRAS, 202, 59

\bibitem[\protect\citeauthoryear{Cardelli, Clayton, \& Mathis}{1989}]{car89}
Cardelli J.~A., Clayton G.~C., Mathis J.~S., 1989, ApJ, 345, 245

\bibitem[\protect\citeauthoryear{Cohen et al.}{2003}]{coh03a}
Cohen M., Megeath S.~T., Hammersley P.~L., Mart{\'{\i}}n-Luis F., Stauffer, J., 2003, AJ, 125, 2645

\bibitem[\protect\citeauthoryear{Cohen, Wheaton \& Megeath}{2003}]{coh03b}
Cohen M., Wheaton W.~A., Megeath S.~T., 2003, AJ, 126, 1090

\bibitem[\protect\citeauthoryear{Cutri et al.}{2003}]{cut03}
Cutri, R. M., Skrutskie, M. F., van Dyk, S., et al. 2003, 2MASS All Sky Catalog
of Point Sources (Pasadena, CA: NASA/IPAC)


\bibitem[\protect\citeauthoryear{Derekas et al.}{2006}]{der06}
Derekas A., Kiss L. L., Bedding T. R., Kjeldsen H., Lah P., Szab\'o G. M., 2006, ApJ, 650, L55 

\bibitem[\protect\citeauthoryear{Fraser et al.}{2008}]{fra08}
Fraser O. J., Hawley S. L., Cook K. H., 2008, AJ, 136, 1242


\bibitem[\protect\citeauthoryear{Gordon et al.}{2011}]{gor11} 
Gordon K.~D., Meixner M., Meade M.~R., et al. 2011, AJ, 142, 102 


\bibitem[\protect\citeauthoryear{Gustafsson et al.}{2008}]{gus08}
Gustafsson B., Edvardsson B., Eriksson K., et al. 2008, A\&A, 486, 951
	


\bibitem[\protect\citeauthoryear{Hinkle et al.}{2002}]{hin02} 
Hinkle K. H., Lebzelter T.,  Joyce R. R., Fekel F. C., 2002, AJ, 123, 1002

\bibitem[\protect\citeauthoryear{Houdashelt et al.}{2000}]{hou00}
Houdashelt M.~L., Bell R.~A., Sweigart A.~V., Wing R.~F., 2000, AJ, 119, 1424 

\bibitem[\protect\citeauthoryear{IRAC Data Handbook}{2006}]{irac06}
IRAC Data Handbook, version 3.0, 2006 January 20

\bibitem[\protect\citeauthoryear{Ita et al.}{2018}]{ita18}
Ita Y., Matsunaga N., Tanab\'e T., Nakada Y., Kato D., Nagayama T., Nagashima C., Kurita M., Nakajima Y., Whitelock P.~A., Menzies J.~W., Feast M.~W., Nagata T., Tamura M., Nakaya H., 2018, MNRAS, 481, 4206

\bibitem[\protect\citeauthoryear{Ita et al.}{2004}]{ita04}
Ita Y., Tanab\'e T.,  Matsunaga N.,  Nakajima Y., Nagashima C., 
Nagayama T, Kato D.,  Kurita M.,  Nagata T., Sato S., Tamura M., Nakaya H.,
Nakada Y., 2004, MNRAS, 347, 720

\bibitem[\protect\citeauthoryear{Keller \& Wood}{2006}]{kel06}
Keller S.~C., Wood P.~R., 2006, ApJ, 642, 834

\bibitem[\protect\citeauthoryear{Ku{\v c}inskas et al.}{2005}]{kuc05}
Ku{\v c}inskas A., Hauschildt P.~H., Ludwig H.-G., Brott I., Vansevi{\v c}ius V., Lindegren L., Tanab\'e T., Allard F., 2005A\&A, 442, 281
	
\bibitem[\protect\citeauthoryear{Lomb}{1976}]{lom76}
Lomb N. R., 1976, Ap\&SS, 39, 447L

\bibitem[\protect\citeauthoryear{Marigo, Girardi \& Chiosi}{2003}]{mar03}
Marigo P., Girardi L., Chiosi C., 2003A\&A, 403, 225

\bibitem[\protect\citeauthoryear{Mennickent et al.}{2002}]{men02}
Mennickent, R. E., Pietrzy\'nski, G., Gieren, W., Szewczyk, O., 2002, A\&A, 393, 887

\bibitem[\protect\citeauthoryear{Neilson \& Lester}{2013}]{nei13} 
Neilson H. R., Lester J. B., 2013, A\&A, 554, A98

\bibitem[\protect\citeauthoryear{Nicholls et al.}{2009}]{nic09} 
Nicholls C. P., Wood P. R., Cioni M.-R. L., Soszy\'nski I.,
2009, MNRAS, 399, 2063

\bibitem[\protect\citeauthoryear{Nie et al.}{2012}]{nie12} 
Nie J.D., Wood P. R., Nicholls C. P.,
2012, MNRAS, 423, 2764

\bibitem[\protect\citeauthoryear{Olivier \& Wood}{2003}]{oli03} 
Olivier E. A., Wood P. R., 2003, ApJ, 584, 1035

\bibitem[\protect\citeauthoryear{Saio et al.}{2015}]{saio15} 
Saio H., Wood P.~R., Takayama M., Ita Y., 2015, MNRAS, 452, 3863


\bibitem[\protect\citeauthoryear{Scargle}{1982}]{sca82}
Scargle J. D., 1982, ApJ, 263, 835

\bibitem[\protect\citeauthoryear{Scowcroft et al.}{2016}]{sco16}
Scowcroft V., Freedman W.~L., Madore B.~F., Monson A., Persson S.~E., Rich J., Seibert M., Rigby J.~R., 2016, ApJ, 816, 49S


\bibitem[\protect\citeauthoryear{Smith et al.}{2002}]{smi02}	
Smith B. J., Leisawitz D., Castelaz M. W., Luttermoser D., 2002, AJ, 123, 948

\bibitem[\protect\citeauthoryear{Soker \& Clayton}{1999}]{sok99} 
Soker N., Clayton G. C., 1999, MNRAS, 307, 993

\bibitem[\protect\citeauthoryear{Soszy\'nski}{2007}]{sos07b} 
Soszy\'nski I., 2007, ApJ, 660, 1486

\bibitem[\protect\citeauthoryear{Soszy\'nski et al.}{2007}]{sos07} 
Soszy\'nski I., Dziembowski W.~A., Udalski A., Kubiak M., Szyma\'nski M.~K., Pietrzy\'nski G., Wyrzykowski \L., Szewczyk O., Ulaczyk K., 2007,
AcA, 57, 201

\bibitem[\protect\citeauthoryear{Soszy\'nski \& Udalski}{2014}]{sos14} 
Soszy\'nski I., Udalski A., 2014, ApJ, 788, 13

\bibitem[\protect\citeauthoryear{Soszy\'nski et al.}{2004a}]{sos04} 
Soszy\'nski I., Udalski A., Kubiak M., Szyma\'nski M., Pietrzy\'nski G.,  \.{Z}ebru\'n K.,
Szewczyk O., Wyrzykowski \L.,  2004a, AcA, 54, 129

\bibitem[\protect\citeauthoryear{Soszy\'nski et al.}{2004b}]{sos04_seqE} 
Soszy\'nski I., Udalski A., Kubiak M., Szyma\'nski M.~K., Pietrzy\'nski G., \.{Z}ebru\'n K.,
Szewczyk O., Wyrzykowski \L., Dziembowski W.~A., 2004b, AcA, 54, 347

\bibitem[\protect\citeauthoryear{Soszy\'nski et al.}{2011}]{sos11} 
Soszy\'nski I., Udalski A., Szyma\'nski M.~K., Kubiak M., Pietrzy\'nski G., Wyrzykowski \L., Ulaczyk K., Poleski R., Koz{\l}owski S., Pietrukowicz P., 2011, AcA, 61, 217

\bibitem[\protect\citeauthoryear{Soszy\'nski \& Wood}{2013}]{sos13b} 
Soszy\'nski I., Wood P. R., 2013, ApJ, 763, 103

\bibitem[\protect\citeauthoryear{Stello et al.}{2014}]{ste14}
Stello D., Compton D.~L., Bedding T.~R., Christensen-Dalsgaard J., Kiss L.~L., Kjeldsen H., Bellamy B., Garc{\'{\i}}a R.~A., Mathur S., 2014, ApJ, 788, L10

\bibitem[\protect\citeauthoryear{Stothers}{2010}]{sto10}
Stothers R. B., 2010, ApJ, 725, 1170

\bibitem[\protect\citeauthoryear{Tabur et al.}{2010}]{tab10}
Tabur V., Bedding T. R., Kiss L. L., Giles T., Derekas A., Moon T. T.,
2010, MNRAS, 409, 777

\bibitem[\protect\citeauthoryear{Takayama, Saio \& Ita}{2013}]{takayama13}
Takayama M., Saio H., Ita Y., 2013, MNRAS, 431, 3189

\bibitem[\protect\citeauthoryear{Takayama, Wood \& Ita}{2015}]{takayama15}
Takayama M., Wood P. R., Ita Y., 2015, MNRAS, 448, 464

\bibitem[\protect\citeauthoryear{Trabucchi et al.}{2017}]{tar17}
Trabucchi M., Wood P. R., Montalb\'an J., Marigo P., Pastorelli G., Girardi L\'eo., 2017, ApJ, 847, 139

\bibitem[\protect\citeauthoryear{Wood}{2015}]{woo15}
Wood P. R., 2015, MNRAS, 448, 3829

\bibitem[\protect\citeauthoryear{Wood et al.}{1999}]{woo99} 
Wood P. R., Alcock C., Allsman R. A., Alves D., Axelrod T. S., Becker A. C., Bennett D. P.,
Cook K. H., Drake A. J., Freeman K. C., et al. 1999, IAU Symp., 191, 151

\bibitem[\protect\citeauthoryear{Wood \& Nicholls}{2009}]{woo09} 
Wood P. R., Nicholls C. P., 2009, ApJ, 707, 573

\bibitem[\protect\citeauthoryear{Wood, Olivier \& Kawaler}{2004}]{woo04} 
Wood P. R., Olivier E. A, Kawaler S. D., 2004, ApJ, 604, 800

\bibitem[\protect\citeauthoryear{Zaritsky et al.}{2002}]{zar02} 
Zaritsky D., Harris J., Thompson I.~B.,  Grebel E.~K., Massey P., 2002, AJ, 123, 855

\end{thebibliography}
\end{document}